\documentclass[final,10pt]{IEEEtran}
\usepackage{lipsum} 

\usepackage{soul}

\usepackage{graphicx}
\usepackage{float}
\usepackage{color}
\usepackage{caption}
\usepackage{subcaption}

\usepackage{amsmath,amssymb,amsthm,mathtools}
\usepackage{bm}
\usepackage{bbm}
\usepackage{mathrsfs}
\usepackage{upgreek}

\usepackage{cite}

\usepackage{algorithm, algpseudocode}

\usepackage[acronym,shortcuts]{glossaries}

\newacronym{5G}{5G}{fifth generation}
\newacronym{6G}{6G}{sixth generation}
\newacronym{AER}{AER}{activity error rate}
\newacronym{AP}{AP}{access point}
\newacronym{AMP}{AMP}{approximate message passing}
\newacronym{AUD}{AUD}{active user detection}
\newacronym{AWGN}{AWGN}{additive white Gaussian noise}
\newacronym{BER}{BER}{bit error rate}
\newacronym{BiGAMP}{BiGAMP}{bilinear generalized approximate message passing}
\newacronym{BiGaBP}{BiGaBP}{bilinear Gaussian belief propagation}
\newacronym{BP}{BP}{belief propagation}
\newacronym{BS}{BS}{base station}
\newacronym{CCU}{CCU}{central computing unit}
\newacronym{CE}{CE}{channel estimation}
\newacronym{CSI}{CSI}{channel state information}
\newacronym{CSIDCO}{CSIDCO}{complex SIDCO}
\newacronym{DoF}{DoF}{degrees of freedom}
\newacronym{eMBB}{eMBB}{enhanced mobile broadband}
\newacronym{FA}{FA}{false alarm}
\newacronym{GaBP}{GaBP}{Gaussian belief propagation}
\newacronym{i.i.d.}{i.i.d.}{independent and identically distributed}
\newacronym{JACDE}{JACDE}{joint activity, channel and data estimation}
\newacronym{JACE}{JACE}{joint activity and channel estimation}
\newacronym{JCDE}{JCDE}{joint channel and data estimation}
\newacronym{MAC}{MAC}{multiple-access channel}
\newacronym{MD}{MD}{miss-detection}
\newacronym{MNS}{MNS}{minimum norm solution}
\newacronym{MIMO}{MIMO}{multiple-input multiple-output}
\newacronym{MMSE}{MMSE}{minimum mean square error}
\newacronym{mMTC}{mMTC}{massive machine type communications}
\newacronym{MMV-AMP}{MMV-AMP}{multiple measurement vector approximate message passing}
\newacronym{MSE}{MSE}{mean square error}
\newacronym{MUD}{MUD}{multi-user detection}
\newacronym{NR}{NR}{new radio}
\newacronym{NMSE}{NMSE}{normalized mean square error}
\newacronym{OFDM}{OFDM}{orthogonal frequency-division multiplexing}
\newacronym{PDF}{PDF}{probability density function}
\newacronym{QP}{QP}{quadratic program}
\newacronym{QPSK}{QPSK}{quadrature phase-shift keying}
\newacronym{SIDCO}{SIDCO}{sequential iterative decorrelation via convex optimization}
\newacronym{SNR}{SNR}{signal-to-noise ratio}
\newacronym{URA}{URA}{unsourced random access}
\newacronym{URLLC}{URLLC}{ultra reliable low latency communications}
\newacronym{ZF}{ZF}{zero-forcing}

\newcommand{\diag}[1]{{\mathrm{diag}}\left(#1\right)}
\newcommand{\E}[2]{\mathbb{E}_{#1}\left[{#2}\right]}
\def\NoNumber#1{{\def\alglinenumber##1{}\State #1}\addtocounter{ALG@line}{-1}}

\theoremstyle{definition}
\newtheorem{definition}{Definition}
\theoremstyle{theorem}
\newtheorem{theorem}{Theorem}

\begin{document}
\newcommand\titlename{Grant-Free Access via Bilinear Inference for Cell-Free MIMO with Low-Coherent Pilots}
\title{\titlename}

\author{
Hiroki~Iimori,~\IEEEmembership{Graduate Student Member,~IEEE,}
Takumi~Takahashi,~\IEEEmembership{Member,~IEEE,}
Koji~Ishibashi,~\IEEEmembership{Member,~IEEE,}
Giuseppe~Thadeu~Freitas~de~Abreu, \IEEEmembership{Senior Member, IEEE,}
and~Wei~Yu,~\IEEEmembership{Fellow,~IEEE,}
\thanks{H. Iimori and G. T. F. Abreu are with the Focus Area Mobility, Department of Computer Science and Electrical Engineering, Jacobs University Bremen, Campus Ring 1, 28759, Bremen, Germany, (email: h.iimori@ieee.org, g.abreu@jacobs-university.de).}
\thanks{T. Takahashi is with Department of Information and Communications Technology, Osaka University, Yamada-oka 2-1, Suita 565-0871, Japan, (email: takahashi@wcs.comm.eng.osaka-u.ac.jp).}
\thanks{K. Ishibashi is with Advanced Wireless \& Communication Research Center (AWCC), The University of Electro-Communications, 1-5-1 Chofugaoka, Chofu-shi, Tokyo 182-8585, Japan, (email: koji@ieee.org).}
\thanks{W. Yu is with Edward S. Rogers Sr. Department of Electrical and Computer Engineering, University of Toronto, Toronto, ON M5S 3G4, Canada, (email: weiyu@comm.utoronto.ca).}
\vspace{-2ex}}

\markboth{Journal of \LaTeX\ Class Files,~Vol.~, No.~, August~2020}%
{H.~Iimori \MakeLowercase{\textit{et al.}}: \titlename}

\maketitle

\begin{abstract}
We propose a novel \ac{JACDE} scheme for cell-free \ac{MIMO} systems compliant with \ac{5G} \ac{NR} \ac{OFDM} signaling.
The contribution aims to allow significant overhead reduction of cell-free \ac{MIMO} systems by enabling grant-free access, while maintaining moderate throughput per user.
To that end, we extend the conventional \ac{MIMO} \ac{OFDM} protocol so as to incorporate activity detection capability without resorting to spreading informative data symbols, in contrast with related work which typically relies on signal spreading.
Our method leverages a Bayesian message passing scheme  based on Gaussian approximation, which jointly performs \ac{AUD}, \ac{CE}, and \ac{MUD}, incorporating also a well-structured low-coherent pilot design based on frame theory, which mitigates pilot contamination, and finally complemented with a detector empowered by bilinear message passing.
The efficacy of the resulting \ac{JACDE}-based grant-free access scheme without spreading data sequences is demonstrated by simulation results, which are shown to significantly outperform the current state-of-the-art and approach the performance of an idealized (genie-aided) scheme in which user activity and channel coefficients are perfectly known.
\end{abstract}

\begin{IEEEkeywords}
Bilinear inference, cell-free MIMO, grant-free access, frame theoretical signaling, 5G new radio
\end{IEEEkeywords}

\glsresetall

\section{Introduction}
\label{sec:intro}

Multiple-antenna architectures, in particular massive \ac{MIMO} and its extensions, will continue to be one of essential technologies in \ac{5G} and future \ac{6G} networks, in order to satisfy the heterogeneous requirements raised by \ac{mMTC}, \ac{eMBB}, \ac{URLLC}, their various combinations, and the ever-growing demand for higher data rates and user capacities \cite{BockelmannCM2016, IimoriTWC19, LarssonMIMOmagagine, ngo2016cellfree,  AndreiAccess19, SanguinettiGLOBECOM2018, IimoriWCL20, RanganRappaport2014, RappaportTAP2017, IimoriAsilomar2019}.
Besides, massive \ac{MIMO} technology is considered an enabler of not only high throughput communications but also massive connectivity due to the significant amount of the spatial \acp{DoF} it provides, which can be exploited to solve inherent problems such as \ac{MUD}, \ac{CE}, and \ac{AUD} in uplink scenarios, among others.

This article aims to tackle the challenging task of non-coherent \ac{JACDE} in a cell-free \ac{MIMO} architecture without spreading the data symbols unlike most of existing works. A key component of the proposed approach is an appropriately designed bilinear message passing algorithm.

As compared to the conventional coherent \ac{MIMO} communication mechanism, where \ac{AUD} and \ac{CE} are sequentially performed based on predetermined reference signals ($e.g.,$ pilot sequences) followed by \ac{MUD} relying on the estimated \ac{CSI}, a main challenge of massive uplink access is the communication overhead for \ac{CSI} acquisition, which scales with the number of potential uplink users in the system due to the need of orthogonal pilot sequences so as to maintain accurate \ac{CSI} knowledge.
It is also worth mentioning that utilizing non-orthogonal pilot sequences for channel estimation, while contributing to reduce overhead, leads to severe \ac{MUD} performance deterioration due to the rank-deficient ($i.e.,$ underdetermined) conditions typically faced, even under the assumption that perfect \ac{AUD} is available at the receiver.
In addition, in massive \ac{MIMO} settings, excessive piloting might exceed channel coherence time, particular in the case of fast fading environments, which makes non-orthogonal pilots necessity.

One of the emerging approaches aimed at tackling this issue is \ac{JCDE} which takes advantage of estimated data symbols as soft pilot sequences, while exploiting their pseudo-orthogonality to improve system performance and efficiency.
In particular, the \ac{BiGAMP} scheme proposed in \cite{ParkerTSP14} has been considered a key ingredient to solve such a detection problem in wireless systems.
In that scheme, Onsager correction is employed to decouple the self-feedback of messages across iterations as is the case with the \ac{AMP}, leading to stable convergence behavior as shown, for instance, in \cite{ParkerTSP14b, Kabashima2016}.

It has, however, been recently shown in the literature \cite{JeremyICASSP15} that the estimation performance of \ac{BiGAMP} severely deteriorates when non-orthogonal pilot sequences are exploited, even if adaptive damping is employed, due to the fact that the derivation of \ac{BiGAMP} relies heavily on the assumption of very large systems, although shortening the pilot sequence is the very aim of the method itself.
In order to circumvent this issue, the authors in \cite{ItoGC20} proposed a novel bilinear message passing algorithm, referred to as \ac{BiGaBP}, with the aim of generalizing \ac{BiGAMP} on the basis of \ac{BP} \cite{KabashimaCDMA03} for robust recovery subject to non-orthogonal piloting.
Despite the aforementioned progresses, many existing works including the ones mentioned above, focus only on joint \ac{CE} and \ac{MUD} while assuming that perfect \ac{AUD} is available at the receiver.

One of the solutions to the \ac{AUD} problem is grant-free random access \cite{AminGlobecom17,LiangSPM18}, which has been intensively investigated in the last few years and can be categorized as a variation of \ac{JACE} or \ac{JACDE}, with Bayesian receiver design components.
In context of Bayesian approaches, Bayesian \ac{JACE} can be seen as a non-orthogonal pilot-based random access protocol in which active users simultaneously transmit their unique spreading signatures to their \acp{BS}, and the \ac{BS} employs a massage passing algorithm -- $e.g.$ \ac{MMV-AMP} -- as receiver, with the aim of detecting user activity patterns and their corresponding channel responses, while taking advantage of the time-sparsity resulting from the activity patterns.
As for Bayesian \ac{JACDE} schemes, most of the existing works on that approach, $e.g.$ \cite{YangWCL18, HaraAccess19, ShuchaoTWC2020, WeijieTCom20,YuanyuanTVT20}, are extensions of the aforementioned Bayesian \ac{JACE} in which spreading data sequences generated by multiplying data symbols with their unique spreading signatures are transmitted, while leveraging a similar receiver design as that of Bayesian \ac{JACE} methods.

There is also another approach to \ac{AUD} that takes advantage of the sample covariance matrix constructed from the large number of antennas at the receiver.
This covariance-based method has also attracted attention due to its applicability to \ac{URA}, where  \ac{JACDE} can be achieved by letting active users transmit a codeword sequence selected from a common predetermined codebook over a given time slot. 
To elaborate, it has been shown in \cite{AlexanderArxiv19} that the covariance-based approach is able to accommodate a larger number of active users, while using limited per-user wireless resources \footnote{In \cite{AlexanderArxiv19, CaireAsilomar2019}, for instance, $96$ bits per user are sent by exploiting $3200$ symbol lengths, which correspond to approximately $11 \sim 22$ \ac{OFDM} frames in \ac{5G} \ac{NR} setups with a sub-GHz carrier frequency. In other words, as a \ac{5G} \ac{NR} \ac{OFDM} frame is designed to amount to $10$ [ms], $96$ bits are delivered with hundreds of milliseconds in this setup.} due to the nature of index-type modulation based on spread codewords, which is therefore suited to low-rate super \ac{mMTC} scenarios.

Besides the above, a fundamental challenge from a system point of view is the spatial correlation of the massive \ac{MIMO} channel, which has been argued in \cite{LarssonCommMag14} to be a limiting factor of centralized massive \ac{MIMO}, although most of the existing work in the area, including \cite{JeremyICASSP15, ItoGC20, YangWCL18, ZhilinTWC19,LiangTSP18,HaraAccess19,XiaodanTSP20}, makes use of the assumption that channels are subjected to ideal (uncorrelated) Rayleigh fading.
In order to iron out this issue, the cell-free massive \ac{MIMO} concept -- studied $e.g.$ in \cite{NgoTWC17, EmilTCom20}, which virtually configures a massive \ac{MIMO} setup by spatially distributing \acp{AP} connected through wired fronthaul links to a common \ac{CCU} -- has recently emerged, offering an architecture capable of decorrelating the spatial dependence between \acp{AP}, leading to an ideal independently distributed channel structure.

In light of the above, the main focus of this article is to incorporate virtues of the aforementioned approaches, proposing a novel grant-free \ac{JACDE} algorithm for non-orthogonal massive random access in cell-free \ac{MIMO} systems with \emph{non}-spread data streams, which, to the best of our knowledge, has not been presented yet in the literature.

\vspace{-2ex}
\subsection{Related Work}
\vspace{-1ex}

As described above, there is a variety of approaches to solve  \ac{MUD}, \ac{CE}, and \ac{AUD} in uplink \ac{MIMO} systems, which for the sake of readability are categorized into two distinct approaches detailed as follows.

First, there are the approaches exploiting bilinear inference, well-known methods including the convex relaxation methods \cite{Kuybeda2013}, the non-convex successive over-relaxation methods \cite{Wen2012}, the variational-Bayes methods \cite{Babacan2012}, and the AMP methods, among others.
In \cite{ParkerTSP14, ParkerTSP14b}, a unified AMP-based approach to matrix completion, robust principle component analysis (PRCA), and dictionary learning was proposed, and the resulting \ac{BiGAMP} was empirically shown in \cite{ParkerTSP14b, Kabashima2016} to be competitive in terms of phase transition and computation complexity.
In the context of self-calibration and matrix compressed sensing, parametric \ac{BiGAMP} (PBiGAMP) was proposed in \cite{Parker2016} and found to yield improved phase transitions in comparison with the aforementioned counterparts.

In the attempt to overcome AMP's vulnerability against measurement correlation \cite{Bayati2011, Bayati2015}, the vector AMP (VAMP) \cite{Rangan2019}, orthogonal AMP (OAMP) \cite{Opper2005} and other iterative detectors based on the expectation propagation (EP) framework \cite{Minka2013, KeigoSPAWC17} have been proposed, which handle a class of unitary-invariant measurement matrices.
These algorithms require, however, matrix inversion operations unlike the original AMP approach.
A rigorous analysis of the convergence property of this approach was presented in \cite{Rangan2019, KeigoSPAWC17}, and potential connection among different methods was investigated in \cite{Meng2018, Zou2018, Liu2019}, with the extension to the bilinear inference method proposed in \cite{Fletcher2019, Meng2019}.
Also, it is worth noting that the \ac{GaBP} \cite{KabashimaCDMA03} approach can be interpreted as the origin of the aforementioned message passing rules.

%
%
%

In the context of grant-free massive random access, it has been shown theoretically in \cite{PolyanskiyISIT17} that non-orthogonal access schemes with random coding not only  outperform classical \ac{MAC} protocols such as coded ALOHA, but also have the potential to nearly achieve theoretical limits in terms of the user capacity.
Motivated by the above, various authors \cite{CaireAsilomar2019, VamsiArxiv18, AlexanderArxiv19, JialinICASSP20, YongpengWC20, Andrea6GSum20} studied random coding schemes and/or its covariance-based receiver designs for \ac{URA} in grant-free \ac{MIMO} systems with massive numbers of potential users.
In this line of work, the massive \ac{MIMO} \ac{JACE} problem was considered in \cite{ZhilinICC19, ZhilinAsilomar19} with a similar receiver design based on the sample covariance approach.
Finally, as for the Bayesian approach, the authors in \cite{LiangTSP18, ZhilinTSP18, MalongTSP20, XiaodanTSP19} have investigated \ac{JACE} for massive random access with Bayesian compressed sensing receivers, which was also extended to \ac{JACDE} in \cite{YangWCL18, HaraAccess19, ShuchaoTWC2020, SenelGlobecom17}, where informative bits are embedded into spreading codewords. 
And since many existing works, including the ones mentioned above, considered the conventional centralized massive \ac{MIMO} architecture, which in practice might suffer from spatial correlation, a more recent contribution \cite{ShuaifeiArxiv2020} has tackled this issue by studying a structured massive access scheme for \ac{JCDE} in a cell-free massive \ac{MIMO} setting, while assuming perfect activity detection and a grant-based architecture.

In summary, it can be said that the majority of contributions addressing \ac{JACDE} in \ac{MIMO} uplink channels can be categorized as either covariance-based receivers for the grant-free \ac{URA}, or Bayesian receivers with informative data symbols embedded into spreading sequences.

\subsection{Contributions}

Given the above, the contributions of the article are summarized as follows.

\begin{itemize}
\item {\bf Feasibility of non-spread \ac{JACDE}}: In contrast to most of related literature \cite{YangWCL18, HaraAccess19, ShuchaoTWC2020, WeijieTCom20}, which addresses \ac{AUD}, \ac{CE}, and \ac{MUD} jointly in a grant-free fashion but which on the other hand require spreading data sequences, we demonstrate the feasibility of grant-free \ac{JACDE} \ul{without spreading data sequences} drawn from a predetermined constellation as is the case for the conventional coherent \ac{MIMO}-\ac{OFDM} systems.
\item {\bf Cell-free assisted grant-free access}: A potential advantage of the cell-free architecture is that helps resolve spatial correlation problems of massive \ac{MIMO}. To the best of our knowledge, however, no grant-free design for cell-free massive \ac{MIMO} scheme without spreading data has been proposed yet. In this article, we extend previous works such as \cite{ZhilinTWC19, JiayiAccess18,ShuaifeiArxiv2020} and offer a \ul{cell-free assisted grant-free access} scheme. 
\item {\bf Frame-theoretic pilot design}: Although much of existing literature relies on Gaussian pilot designs, its mutual coherence is limited unless the large-system limit condition is satisfied. In other words, such a design policy is not suited for a non-orthogonal short pilot aided system, which is the very aim of the article. To this end, a \ul{frame-theoretic non-orthogonal pilot design} is introduced, which asymptotically approaches the theoretical lower bound.
\item {\bf Bilinear inference for \ac{JACDE}}: In order to enable the above, a \underline{novel \ac{JACDE} algorithm, dubbed} \underline{activity-aware \ac{BiGaBP}}, is presented here, in which bilinear inference, message passing rules, Gaussian approximation, and a new belief scaling technique that forges resilience of the derived messages, are combined.
\end{itemize}
 
Simulation results are offered to support the claims above, which is based on the IEEE \ac{5G} \ac{NR} configurations.  
We emphasize that, to the best of our knowledge, a \ac{JACDE} mechanism for cell-free grant-free \ac{MIMO} systems without spreading data symbols, which is the key contribution of this article, has not yet been proposed.

\emph{Notation}: In the remainder of the article, the following notation will be employed.
Sets of numbers in the real, complex, and Hilbert spaces will be denoted by $\mathbb{R}$, $\mathbb{C}$, and $\mathbb{H}$, respectively.
The transpose, transpose conjugate, and hermitian adjoint operators, correspondingly in the real, complex, and Hilbert spaces will be respectively expressed as $\cdot^{\rm T}$, $\cdot^{\rm H}$, and $\cdot^{\rm H}$.
The multivariate circular symmetric complex Gaussian distribution with mean $\bm{\mu}$ and covariance $\bm{\Sigma}$ will be denoted by $\mathcal{CN}\left(\bm{\mu},\bm{\Sigma}\right)$, while $\mathcal{N}\left(\bm{\mu},\bm{\Sigma}\right)$ will denote its real-domain counterpart.
Finally, $\|\cdot\|_p$ will be used to denote the $p$-norm with $p\in\{2,\infty\}$, while $\langle\cdot,\cdot\rangle$ is the inner product operator.

\section{System Model}

\begin{figure}[t!]
\centering
\begin{subfigure}[h]{\columnwidth}
\centering
\includegraphics[width=\columnwidth]{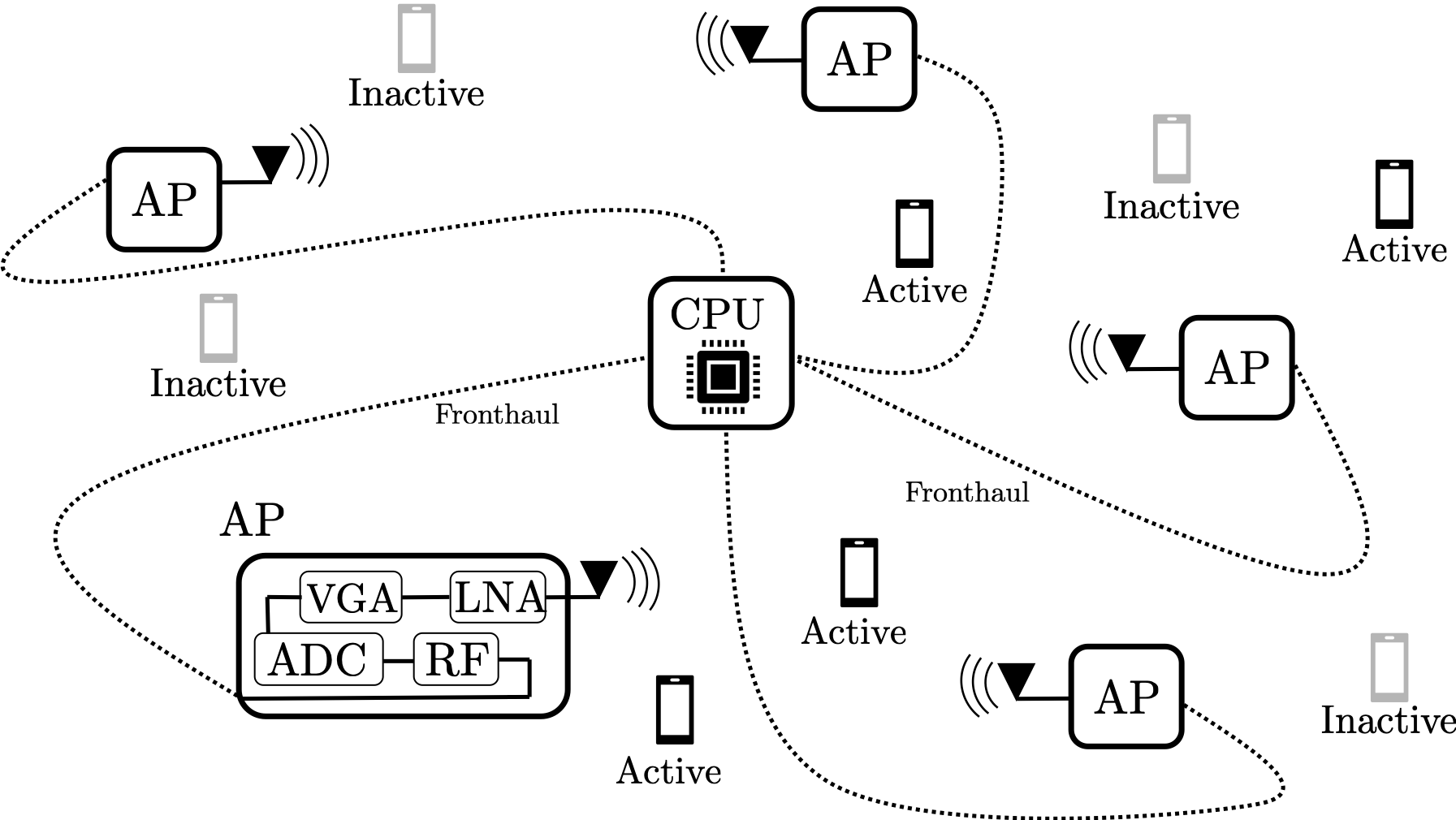}
\caption{A model illustration of cell-free \ac{MIMO} systems with distributed single-antenna \acp{AP} serving uplink users which access the system in a grant-free basis.}
\label{fig:system_model}
\vspace{1.5ex}
\end{subfigure}
\begin{subfigure}[h]{\columnwidth}
\centering
\includegraphics[width=\columnwidth]{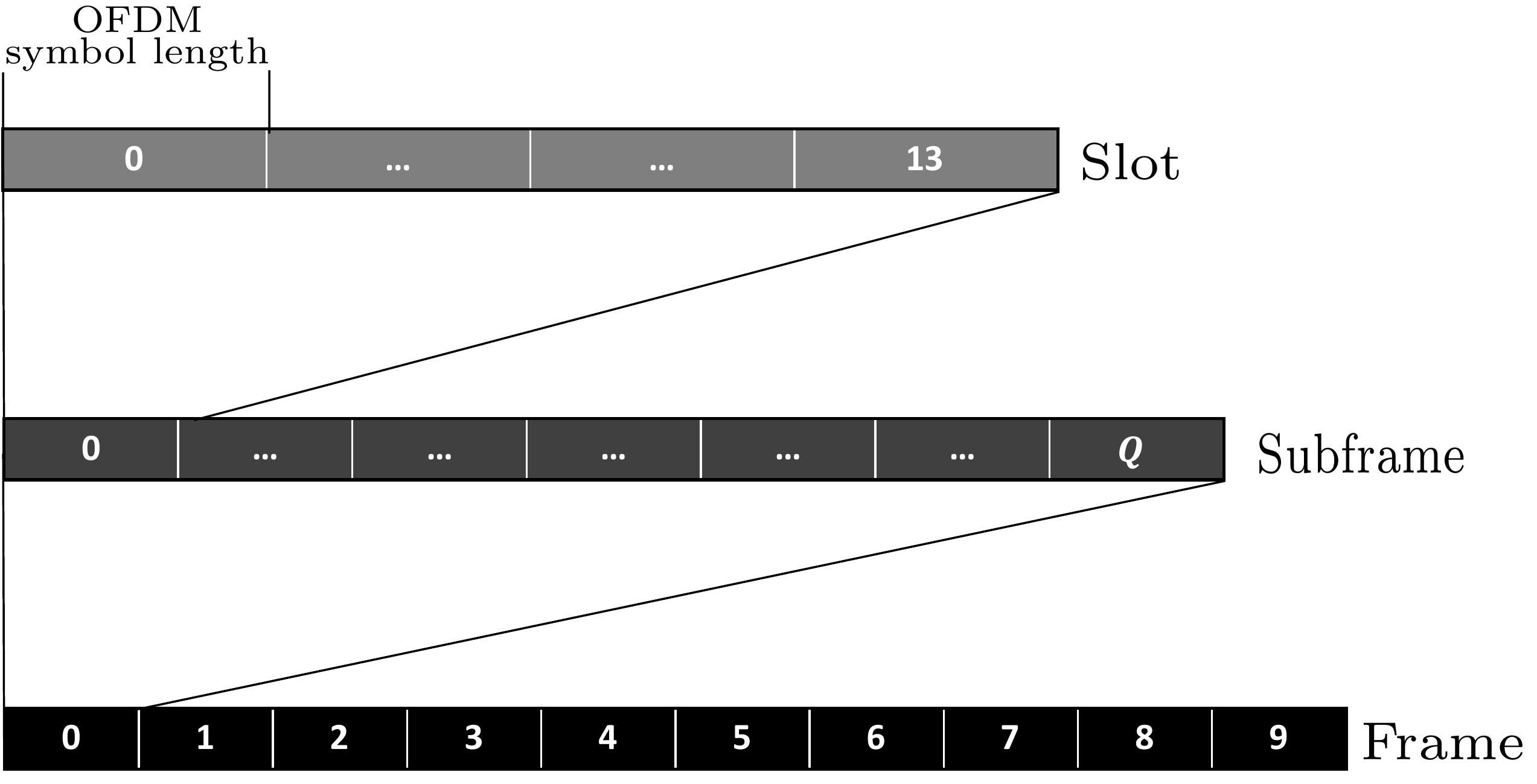}
\caption{Radio frame structure in \ac{5G} \ac{NR} \cite{3GPPTS38101}.}
\label{fig:frame_structure}
\end{subfigure}
\caption{System and Signal Model.}
\label{fig:Model}
\end{figure}

Consider a cell-free large \ac{MIMO} system composed of $N$ spatially distributed single-antenna \acp{AP} connected by wired fronthaul links to a common high-performance \ac{CCU}, serving $M$ synchronized single-antenna users in a grant-free fashion, as depicted in Figure \ref{fig:system_model}. 
Due to the dynamic nature of grant-free systems, it is assumed that a fraction of the total $M$ single-antenna users become active within a given \ac{5G} \ac{NR} \ac{OFDM} symbol frame \cite{3GPPTS38101}, whereas the rest of uplink users remain silence during that period.
Recalling the frame structure as \ac{5G} \ac{NR} as shown in Figure \ref{fig:frame_structure}, in which an \ac{OFDM} frame is composed of $10$ subframes and each subframe consists of $Q$ slots, each comprising $14$ \ac{OFDM} symbols, one may take advantage of a part of the radio frame structure as reference signals and the rest as a data stream.
We also remark that the number of slots within a certain subframe ($i.e.,$ $Q$) depends on the subcarrier spacing employed in a system.

Given the above, let us define $K$ being the total number of discrete time indexes within an \ac{OFDM} frame and $\mathscr{C}\triangleq \{c_1,c_2,\ldots,c_{2^b}\}$ representing a set of constellation points with $b$ denoting the number of bits per symbol.
Introducing the user index set $\mathscr{M}\triangleq\{1,2,\ldots,M\}$ and a set of active users $\mathscr{A}$, the received signal vector $\bm{y}_k \in\mathbb{C}^{N\times 1}$ at $k$-th time index with $k\in\{1,2,\ldots,K\}$ can be written as
\begin{equation}
\label{eq:received_vector}
\bm{y}_k = \bm{H}\bm{x}_k + \bm{w}_k.
\end{equation}

In the system model equation above, $\bm{x}_k\in\mathbb{C}^{M\times 1}$ denotes the transmitted signal vector with non-zero element at index $a\in\mathscr{A}$ and zero otherwise; $\bm{w}_k\in\mathbb{C}^{N\times 1}$ denotes the \ac{AWGN} vector distributed as $\bm{w}_k\sim\mathcal{CN}_N\left(\mathbf{0},N_0\mathbf{I}_N\right)$; and $\bm{H}\in\mathbb{C}^{N\times M}$ denotes the flat fading communication channel matrix, whose element at $n$-th row and $m$-th column follows an \ac{i.i.d.} circularly symmetric complex Gaussian distribution with zero mean and variance $\gamma_{nm}\triangleq 10^{-\tfrac{\beta_{nm}}{10}}$; and $\beta_{nm}$ follows the 3GPP urban microcell model \cite{EmilTWC20}, namely,
\begin{equation}
\beta_{nm} [dB] = 30.5 + 36.7\log_{10}\left(d_{nm}\right) + \mathcal{N}\left(0,4^2\right),
\end{equation} 
with $d_{nm}$ denoting the distance between the $n$-th \ac{AP} and $m$-th user, while assuming that the height of the \ac{AP} and the user are $10$ [m] and $1.65$ [m], respectively.

Assuming that the user activity remains consistent within an \ac{OFDM} frame, we can concatenate $K$ consecutive symbol transmissions into the transmitted signal matrix $\bm{X}\triangleq\left[\bm{x}_1,\bm{x}_2,\ldots,\bm{x}_K\right]$.
In turn, the row sparse nature of $\bm{X}$ translates to a column-sparsity in the channel matrix $\bm{H}$.
To be more specific, the $m$-th column of the channel matrix ($i.e.,$ $\bm{h}_m$) can be modeled as a multivariate random variable following the Bernoulli-Gaussian distribution, that is,
\begin{equation}
\bm{h}_m \sim (1-\lambda)\delta(\bm{h}_m) + \lambda \mathcal{CN}_N\left(\mathbf{0},\bm{\Gamma}_m\right),
\end{equation}
where $\lambda$ is the activity factor, $\delta(\bm{h}_m)$ denotes the Dirac delta function that takes value $1$ if and only if $\bm{h}_m = \mathbf{0}$ and $0$ otherwise, and $\bm{\Gamma}_m\triangleq\diag{[\gamma_{1m},\gamma_{2m},\ldots,\gamma_{Nm}]}$ is the covariance matrix of the channel.

In light of the above, the received signal matrix $\bm{Y}\in\mathbb{C}^{N\times K}$ concatenating the received signal vectors given in equation \eqref{eq:received_vector} over $K$ successive time indices, can be readily expressed as
\begin{equation}\label{eq:RXsignals}
\bm{Y} = \bm{H}\bm{X} + \bm{W},
\end{equation}
where $\bm{Y}\triangleq \left[\bm{y}_1,\bm{y}_2,\ldots,\bm{y}_K\right]$ and $\bm{W}\triangleq \left[\bm{w}_1,\bm{w}_2,\ldots,\bm{w}_K\right]$.

Without loss of generality, in order to explicitly express the the pilot and data sequences within the transmit and received signal, let us define, 
\begin{subequations}
\begin{eqnarray}
\bm{Y} &\!\!\!\!\!\triangleq\!\!\!\!& [\bm{Y}_p, \bm{Y}_d],\: \bm{Y}_p\in\mathbb{C}^{N\times K_p}, \:  \bm{Y}_d\in\mathbb{C}^{N\times K_d},\\
\bm{X} &\!\!\!\!\!\triangleq\!\!\!\!& [\bm{X}_p, \bm{X}_d],\: \bm{X}_p\in\mathbb{C}^{M\times K_p}, \:  \bm{X}_d\in\mathbb{C}^{M\times K_d}\!\!,
\end{eqnarray}
\end{subequations}
with $\bm{\cdot}_p$ and $\bm{\cdot}_d$ corresponding to pilot and data signal blocks, respectively, $K_p + K_d = K$ with $K_p\ll K_d < K$, and each element of $\bm{X}_d$ assumed to be drawn from the constellation set $\mathscr{C}$ in a similar way to conventional coherent \ac{OFDM} systems.

We remark that unlike most grant free systems found in literature, $e.g.$ \cite{ShuchaoTWC2020, HaraAccess19, YangWCL18,ByeongTVT18,DingTWC19,MalongTSP20}, in which informative data is transmitted in the form of a spread sequence, equation \eqref{eq:RXsignals} can be seen as an activity-aware variant of the conventional \ac{MIMO}-\ac{OFDM} systems, which aims at both grant-free random access and pilot length reduction enabled by jointly extrapolating user activity, channel state, and data streams.
These features make our approach suited to grant-free cell-free systems, which better meet the demand of higher spectrum efficiency per user than methods previously proposed, including those aforementioned.

\subsection{Pilot Sequence Design}
\label{sec:pilot}

In this section we apply a frame-theoretic approach to effectively design a structured pilot matrix for non-orthogonal transmission, aiming at efficiently reducing the pilot length $K_p$ while preserving the linear independence between vectors in the pilot matrix $\bm{X}_p$ as much as possible. 
To this end, assuming a non-orthogonal representation of the pilot matrix ($i.e.,$ $K_p<M$), let us first define the mutual coherence in the following as a measure of the similarity between non-orthogonal bases. 

\begin{definition}[Mutual Coherence]
\quad\\
Let $\bm{F}\triangleq \left[\bm{f}_1,\bm{f}_2,\ldots,\bm{f}_L\right]\in\mathbb{H}^{J\times L}$ be a \emph{frame} matrix over a Hilbert space $\mathbb{H}^{J\times L}$, comprising of frame vectors $\bm{f}_\ell\in\mathbb{H}^{J\times 1}$, with $\ell\in\{1,2\ldots,L\}$ and $J<L$.
The mutual coherence of $\bm{F}$ is given by
\vspace{-0.5ex}
\begin{equation}
\mu(\bm{F}) \triangleq {\mathop {\mathrm{max}} \limits_{\ell\neq\ell^\prime}} \frac{|\langle\bm{f}_\ell,\bm{f}_{\ell^\prime}\rangle|}{\|\bm{f}_\ell\|_2\|\bm{f}_{\ell^\prime}\|_2}, \forall \{\ell,\ell^\prime\}\in\{1,2\ldots,L\},
\vspace{-0.5ex}
\end{equation}
which in case of an equal-norm frame, resumes to
\begin{equation}
\vspace{-0.5ex}
\mu(\bm{F}) \triangleq {\mathop {\mathrm{max}} \limits_{\ell\neq\ell^\prime}} |\langle\bm{f}_\ell,\bm{f}_{\ell^\prime}\rangle|, \forall \{\ell,\ell^\prime\}\in\{1,2\ldots,L\}.
\vspace{-0.5ex}
\end{equation} 
\end{definition}

Given the above, one may readily notice that the mutual coherence of a frame matrix $\bm{F}$ is equivalent to the maximum non-diagonal element of its gram matrix $\bm{G}\triangleq \bm{F}^{\rm H}\bm{F}$, which is known to be lower-bounded by the Welch bound for $L\leq J^2$ \cite{Strohmer2003}, that is,
\vspace{-0.5ex}
\begin{equation}
\mu(\bm{F}) \geq\sqrt{\tfrac{L-J}{J(L-1)}},
\vspace{-0.5ex}
\end{equation}
indicating that the Welch bound is a lower-bound on the mutual coherence of a frame.

Furthermore, equally spreading non-orthogonal bases over a certain Hilbert space is necessary for a design of the pilot sequences especially in grant-free systems so as to preserve an analogous tight  energy representation in terms of the fairness among  sporadic uplink users.
To this end, another key property of a frame is introduced in the following.

\begin{definition}[Tightness]
\quad\\
By means of the Rayleigh-Ritz Theorem, a frame matrix $\bm{F}$ possesses the following inequalities.
\begin{equation}
\alpha \|\bm{a}\|^2_2 \leq \| \bm{F}^{\rm H}\bm{a}\|^2_2\leq \beta \|\bm{a}\|^2_2, \forall \bm{a}\in\mathbb{H}^J,
\end{equation}
where $0<\alpha\leq\beta<\infty$ and $\bm{F}$ is called \emph{tight} iif $\alpha=\beta$.
\end{definition}

In light of the above, our goal is to find a frame that minimizes the mutual coherence in order to distinguish each pilot sequence even under severe non-orthogonal scenarios ($i.e.,$ $K_p\ll M$) while maintaining the tightness for the sake of fair transmission in terms of energy consumption.
Although it has been recently shown in \cite{HassibiTSP15} that a group-theoretic tight frame construction approach can indeed achieve near-Welch-bound performance, the algorithm proposed thereby has limitations on the size of a frame matrix constructable via the cyclic-group based method, $e.g.,$ the number of frame vectors ($i.e.,$ $L$) needs to be a prime number. 
Since it is desirable to design a frame with any $L$ and $J$ for system-and-user centric communication architectures, we hereafter consider a convex optimization based frame construction method that is capable of dealing with any dimensional pair while minimizing the mutual coherence from an optimization perspective.

Such a low-coherent unit-norm tight frame with arbitrary dimensions can be obtained by taking advantage of an iterative method, referred to as \ac{SIDCO}, which sequentially separates non-orthogonal bases from any equal-norm frame as an initial starting point.
Since the original \ac{SIDCO} proposed in \cite{RusuTSP16} is limited to frames only in the real space $\mathbb{R}$, its extension to the complex space $\mathbb{C}$, dubbed as \ac{CSIDCO}, has been studied in \cite{RusuSP18, AndreiVTC18}, where the strategy to minimize the mutual coherence is to iteratively solve a series of the following convex optimization problem for a given $\tilde{\bm{F}}\in\mathbb{C}^{J\times L}$. \!
\vspace{-2ex}
\begin{subequations}
\label{eq:CSIDCO}
\begin{eqnarray}
&{\mathop {\mathrm{min}} \limits_{\bm{f}_\ell\in\mathbb{C}^{J\times 1}\atop \forall \ell\in\{1,2,\ldots,L\}}}&\: \|\tilde{\bm{F}}^{\rm H}_\ell \bm{f}_\ell\|_\infty\\[-1ex]
&\mathrm{s. t.}&\: \|\bm{f}_\ell - \tilde{\bm{f}}_\ell\|^2_2 \leq T_\ell,\label{eq:T_ball_const_original}
\end{eqnarray}
\end{subequations}
where $\tilde{\bm{F}}_\ell\in\mathbb{C}^{J\times L-1}$ is equivalent to $\bm{F}$ with its $i$-th column pruned and the search region is limited to a multidimensional Euclidean ball with radius $T_\ell$ given by 
\begin{equation}
T_\ell = 1 - {\mathop {\mathrm{max}} \limits_{ \ell^\prime | \ell^\prime\neq\ell}} \frac{|\langle\bm{f}_\ell,\bm{f}_{\ell^\prime}\rangle|^2}{\|\bm{f}_\ell\|^2_2\|\bm{f}_{\ell^\prime}\|^2_2}.
\end{equation}

\begin{figure}[b!]
\centering
\includegraphics[width=\columnwidth]{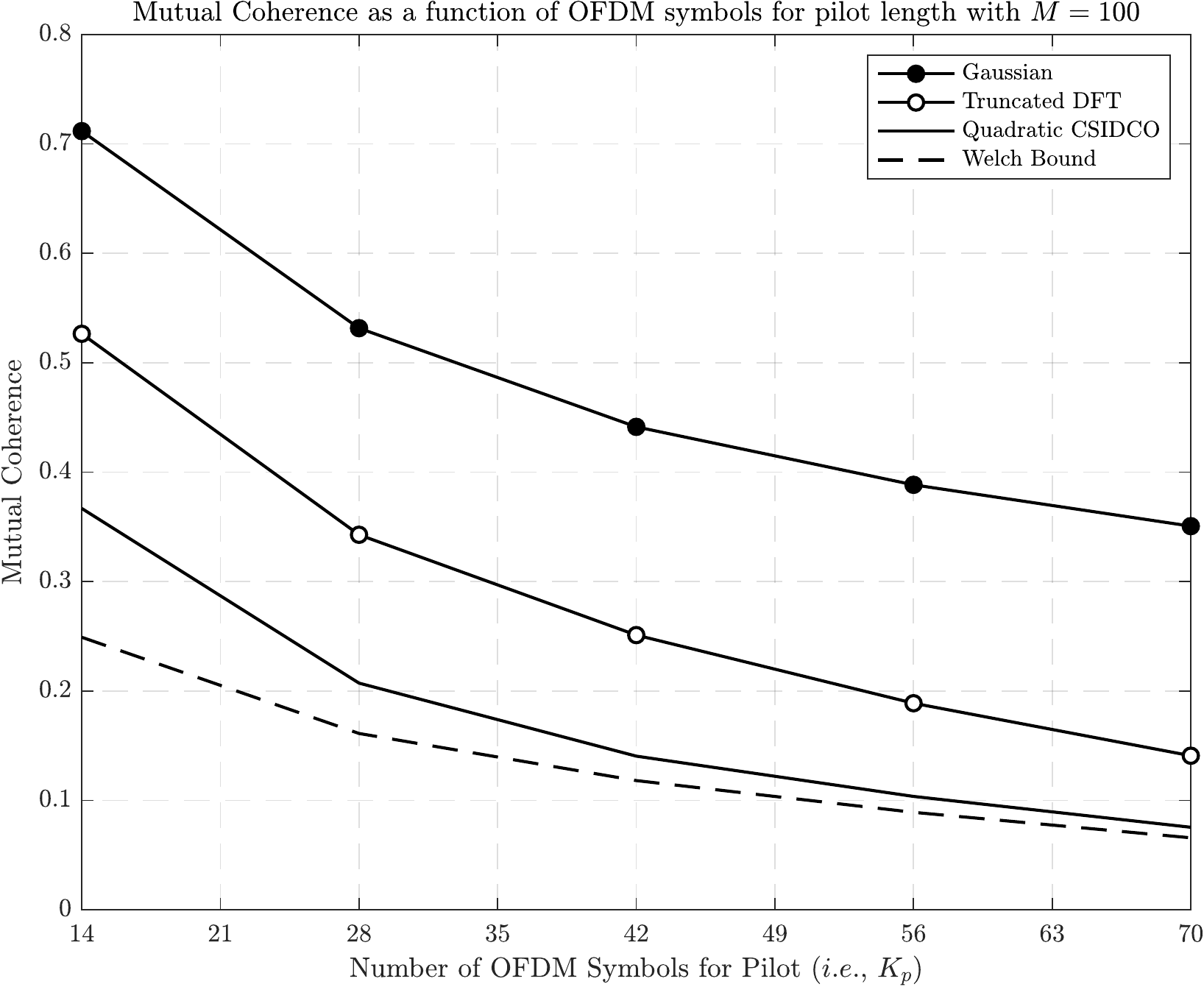}
\caption[]{Mutual coherence comparison as a function of \ac{OFDM} symbols utilized for pilot lengths $K_p$ with $M=100$ uplink users. As benchmarks, we adopt the popular random Gaussian pilot sequence considered in, $e.g.,$, \cite{LiangTSP18,CaireAsilomar2019}, and a randomly truncated discrete Fourier transform matrix.}
\label{fig:MC_comp}
\vspace{-2ex}
\end{figure}

Although the \ac{CSIDCO} reformulation given in equation \eqref{eq:CSIDCO} already follows the disciplined convex programming conventions such that this problem can be easily solved via interior point methods through CVX available in high-level numerical computing programming languages such as MATLAB and Python, the abstraction penalty of such high-level programming languages are too high for real-world communication systems.
Aiming at real-time processing of convex optimization problems, the authors in \cite{MattingleyCVXGEN_SPM12} have developed an automatic low-level code generator for conic programming problems, which solves convex problems with moderate size on the order of microseconds or milliseconds, although it is limited to \ac{QP}-representable convex problems.
In light of the above, for the sake of completeness, equation \eqref{eq:CSIDCO} is transformed into the following quadratic relaxation form.

\begin{theorem}[Quadratic CSIDCO]
\label{theorem:QCSIDCO}
\quad\\
Introducing $\bm{x}_\ell\triangleq\left[\Re\left\{\bm{f}_\ell\right\}; \Im\left\{\bm{f}_\ell\right\}; t_{\ell,R}; t_{\ell,I}\right]\in\mathbb{R}^{2J+2\times 1}$ with slack variables $t_{\ell,R}\in\mathbb{R}_+$ and $t_{\ell,I}\in\mathbb{R}_+$ for all $\ell$, the \ac{CSIDCO} formulation given in equation \eqref{eq:CSIDCO} for a unit-norm low-coherent frame construction can be represented as 
\begin{subequations}
\begin{eqnarray}
\hspace{-5ex}&{\mathop {\mathrm{min}} \limits_{\bm{x}_\ell\atop \forall \ell\in\{1,2,\ldots,L\}}}&\: \bm{x}_\ell^{\rm T}\bm{\Phi}\bm{x}_\ell \\[-0.5ex]
\hspace{-5ex}&\mathrm{s. t.}&\: \bm{A}_{\ell,R,1}\bm{x}_\ell\leq\mathbf{0}\\
\hspace{-5ex}&&\: \bm{A}_{\ell,R,2}\bm{x}_\ell\leq\mathbf{0}\\
\hspace{-5ex}&&\: \bm{A}_{\ell,I,1}\bm{x}_\ell\leq\mathbf{0}\\
\hspace{-5ex}&&\: \bm{A}_{\ell,I,2}\bm{x}_\ell\leq\mathbf{0}\\
\hspace{-5ex}&&\: \bm{x}_\ell^{\rm T}\bm{\Xi}\bm{x}_\ell - 2\bm{b}^{\rm T}_\ell\bm{x}_\ell + 1 - T^2_\ell \leq 0,\label{eq:T_ball_const}
\end{eqnarray}
\end{subequations}
where $\bm{\Phi}\triangleq \begin{bmatrix}\mathbf{0}_{2J} & \mathbf{0}_{2J\times 2}\\\mathbf{0}_{2\times 2J} & \mathbf{I}_{2}\end{bmatrix}$, $\bm{\Xi}\triangleq \begin{bmatrix}\mathbf{I}_{2J} & \mathbf{0}_{2J\times 2}\\\mathbf{0}_{2\times 2J} & \mathbf{0}_{2\times 2}\end{bmatrix}$, $\bm{b}_\ell \triangleq \begin{bmatrix}\tilde{\bm{f}}^{\rm T}_\ell\!\!\! & 0 \!\!& 0 \end{bmatrix}^{\rm T}$, and the other coefficient matrices are listed in the Appendix.
\begin{proof}
See Appendix \ref{app:QCSIDCO}
\end{proof}
\end{theorem}

One may argue that the frame matrix constructed via the quadratic \ac{CSIDCO} method described above is not strictly tight due to the fact that tightness is not enforced during the optimization process.
However, the tightening approach proposed in \cite{TroppTIT05} based on the polar decomposition can be applied to the output of the quadratic \ac{CSIDCO}, yielding an arbitrarily-sized low-coherent equal-norm tight frame sufficiently close to the ideal equiangular tight frames which are not suited to practice as they exist only for particular dimensions, which is therefore adopted as pilot sequences in this article.

To illustrate the performance of the quadratic \ac{CSIDCO} method described above especially in \ac{5G} \ac{NR} setups, mutual coherence comparisons of the method against popular pilot construction approaches ($i.e.,$ the random Gaussian and truncated discrete Fourier transform matrices) as a function of the number of \ac{OFDM} symbols leveraged for pilot lengths are shown in Figure \ref{fig:MC_comp}, which demonstrates that in fact the quadratic \ac{CSIDCO} approaches the Welch bound while reducing the correlation between pilot sequences in comparison with the other two approaches, implying capability of sufficiently mitigating the inter-user interference even in highly non-orthogonal scenarios and efficiently decreasing the number of pilot lengths simultaneously.

\begin{figure}[H]
\centering
\includegraphics[width=\columnwidth]{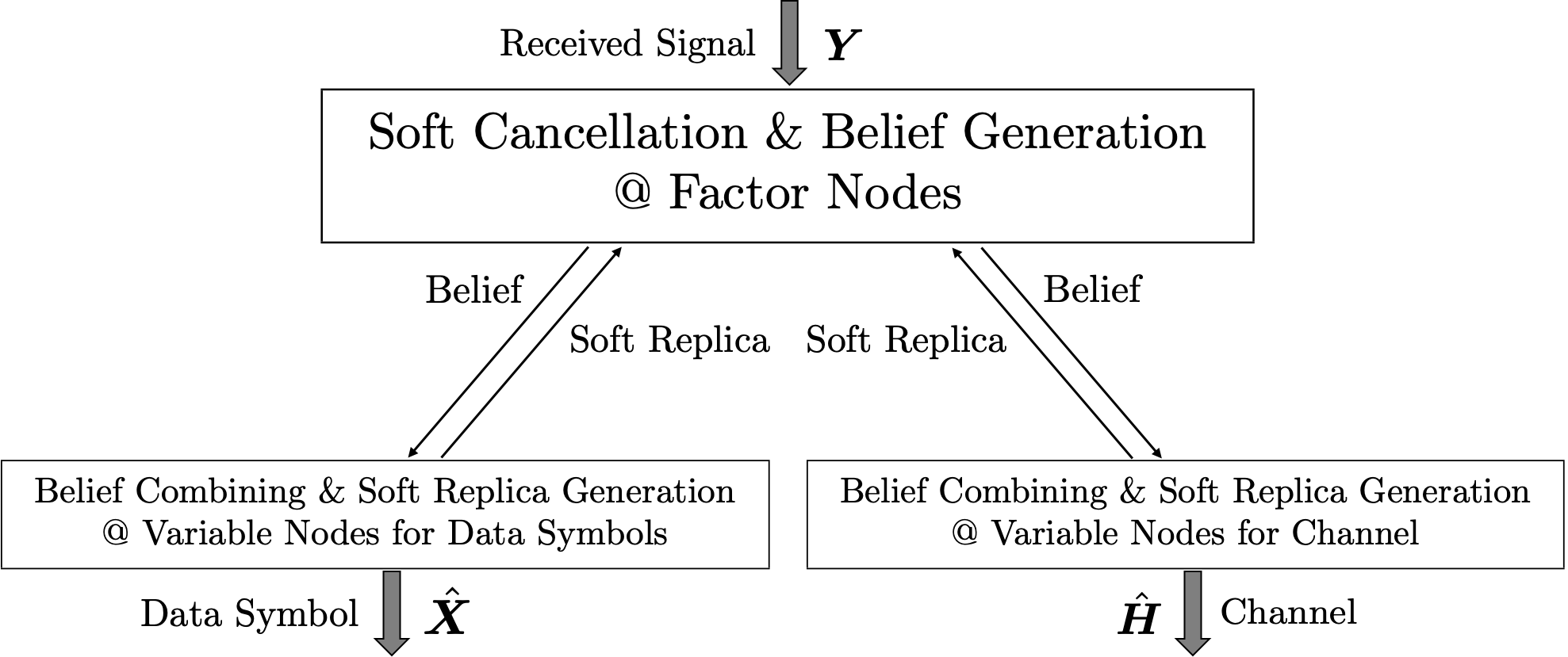}
\caption{Belief generation and combining model.}
\label{fig:schematic_belief_model}
\vspace{1ex}
\centering
\includegraphics[width=0.65\columnwidth]{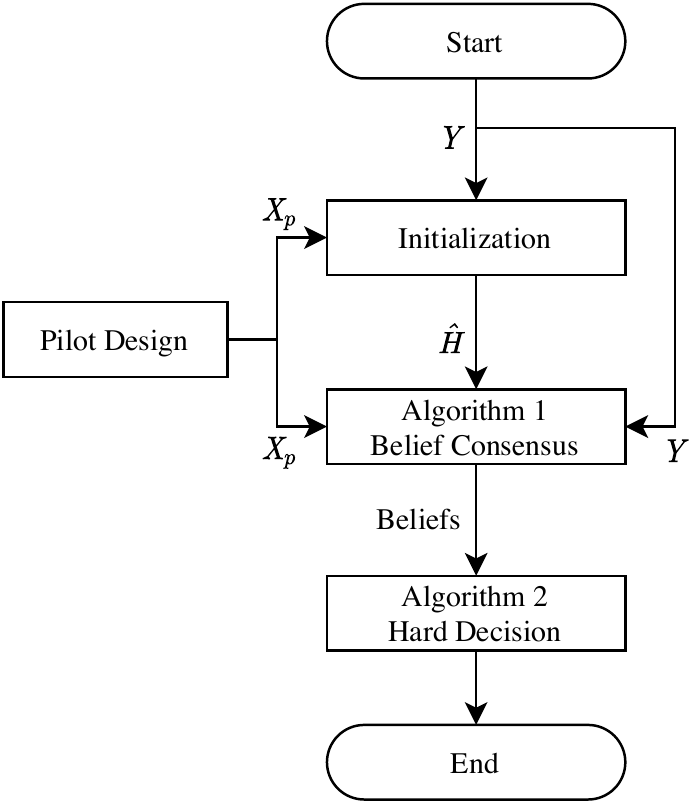}
\caption{Work flow of the proposed detection process.}
\label{fig:work_flow}
\vspace{-3ex}
\end{figure}

\section{Joint Detection via Bilinear Gaussian Belief Propagation}

In this section, we describe a joint activity, channel, and data estimation mechanism via bilinear \ac{GaBP} for large cell-free \ac{MIMO} architectures with \ac{5G} \ac{NR} signaling, whose belief propagation can be modeled as the graph schematized in Figure \ref{fig:schematic_belief_model}.
In order to further clarify the process of the proposed method, a work flow chart of the proposed detection mechanism is also illustrated in Figure \ref{fig:work_flow}, where the proposed detection is split into two algorithms, namely, Algorithm 1): Belief consensus and Algorithm 2): hard decision.

As shown in Figure \ref{fig:work_flow}, the work flow starts with an initialization so as to obtain an initial guess of the channel estimate by leveraging only the pilot sequences, the result of which is however limited in accuracy due to the severely non-orthogonal structure of the pilot matrix as $K_p\ll M$.
Aiming at improving the channel estimation accuracy by jointly detecting the data as well as the activity, the initial channel guess obtained by the initialization process is fed to Algorithm \ref{alg:alg1}, which as illustrated in Figure \ref{fig:schematic_belief_model}, is composed of two different stages; 1) soft interference cancellation and beliefs generation based on tentative soft estimates and 2) combining beliefs and soft estimates generation, described in the next two subsections, respectively.

Followed by Algorithm \ref{alg:alg1}, we proceed with Algorithm \ref{alg:alg2}, where the final hard decision of the data and activity is curried out, which will be described in Section \ref{sec:alg_descript}.

\subsection{Factor nodes}
\label{sec:FN}

Focusing on the received signal element $y_{nk}$ at the $n$-th row and $k$-th column of $\bm{Y}$ with the aim of detecting $x_{mk}$ at the $m$-th row and $k$-th column of $\bm{X}$, the received signal after soft interference cancellation using initial estimates is given by
\vspace{-0.5ex}
\begin{eqnarray}
\label{eq:received_softcanceled}
\tilde{y}_{m,nk} &\!\!\triangleq\!\!& y_{nk} -\hspace{-10ex} \overbrace{\sum^M_{i\neq m}\hat{h}_{k,ni}\hat{x}_{n,ik}}^\text{Inter-user interference cancellation with soft-replicas}\\[-3ex]
&\!\!=\!\!& h_{nm}x_{mk} + \overbrace{\sum^{M}_{i\neq m} (h_{ni}x_{ik} - \hat{h}_{k,ni}\hat{x}_{n,ik})}^
\text{Residual interference} +w_{nk}\nonumber,
\end{eqnarray}
where $\hat{x}_{n,ik}$ and $\hat{h}_{k,ni}$ with $i\in\{1,2,\ldots,M\}$ are tentative estimates of $x_{ik}$ and $h_{ni}$, respectively, generated at variable nodes in the previous iteration, and $w_{nk}$ is the noise element at the $n$-th row and $k$-th column of $\bm{W}$.

Owing to the central limit theorem, the interference-plus-noise component can be approximated as a complex Gaussian random variable under large-system conditions, resulting in the fact that the conditional \ac{PDF} of equation \eqref{eq:received_softcanceled} for given $x_{mk}$ and $h_{nm}$ can be respectively expressed as
\begin{subequations}
\label{eq:Gaussian_belief}
\vspace{-1ex}
\begin{equation}
p_{\tilde{y}_{m,nk}|x_{mk}}(\tilde{y}_{m,nk}|x_{mk}) \propto e^{-\tfrac{|\tilde{y}_{m,nk} - \hat{h}_{k,nm}x_{mk}|^2}{v^x_{m,nk}}},
\vspace{-1ex}
\end{equation}
\vspace{-2ex}
\begin{equation}
p_{\tilde{y}_{m,nk}|h_{nm}}(\tilde{y}_{m,nk}|h_{nm}) \propto e^{-\tfrac{|\tilde{y}_{m,nk} -h_{nm}\hat{x}_{n,mk}|^2}{v^h_{m,nk}}},
\end{equation}
\end{subequations}
with 
\begin{subequations}
\begin{eqnarray}
v^x_{m,nk} &\!\!\!\!\triangleq\!\!\!\!& \sum^{M}_{i\neq m} \left\{|\hat{h}_{k,ni}|^2 \psi^x_{n,ik} + (|\hat{x}_{n,ik}|^2 + \psi^x_{n,ik})\psi^h_{k,ni}\right\}\nonumber\\
&\!\!\!\!\!\!\!\!& + \psi^h_{k,nm} + N_0,\\
v^h_{m,nk} &\!\!\!\!\triangleq\!\!\!\!& \sum^{M}_{i\neq m} \left\{|\hat{h}_{k,ni}|^2 \psi^x_{n,ik} + (|\hat{x}_{n,ik}|^2 + \psi^x_{n,ik})\psi^h_{k,ni}\right\} \nonumber\\
&\!\!\!\!\!\!\!\!& + \gamma_{nm}\psi^x_{n,mk}  + N_0,
\end{eqnarray}
\end{subequations}
where $\psi^x_{n,ik}$ and $\psi^h_{k,ni}$ denote expected error variances corresponding to $\hat{x}_{n,ik}$ and $\hat{h}_{k,ni}$, respectively, which will be explained in detail in the next subsection, and we remark that $\E{}{|x_{mk}|^2}=1$.

\vspace{-1ex}
\subsection{Variable nodes}
\label{sec:VN}

Given the soft interference cancellation and belief generation above, one can combine the Gaussian beliefs given in equation \eqref{eq:Gaussian_belief}, yielding the \ac{PDF} of an extrinsic belief $l^x_{n,mk}$ for $x_{mk}$
\vspace{-1ex}
\begin{eqnarray}
\label{eq:extrinsic_PDF_x}
p_{l^x_{n,mk}|x_{mk}}(l^x_{n,mk}|x_{mk})
\hspace{-4ex}&&=\prod^N_{i\neq n}p_{\tilde{y}_{m,ik}|x_{mk}}(\tilde{y}_{m,ik}|x_{mk})\nonumber\\
&&\propto e^{-\!\frac{|x_{mk} - \hat{r}_{n,mk}|^2}{\psi^r_{n,mk}}},
\end{eqnarray}
with
\begin{subequations}
\begin{eqnarray}
&\hat{r}_{n,mk} \triangleq \psi^r_{n,mk}\sum\limits^N_{i\neq n}\dfrac{\hat{h}^{*}_{k,im}\tilde{y}_{m,ik}}{v^x_{m,ik}}&\\
&\psi^r_{n,mk} \triangleq \Bigg(\sum\limits^N_{i\neq n} \dfrac{|\hat{h}_{k,im}|^2}{v^x_{m,ik}}\Bigg)^{-1}\hspace{-2ex}.&
\end{eqnarray}
\end{subequations}

In turn, since the activity can be expressed as column-sparsity in the channel matrix $\bm{H}$, combining beliefs of the $m$-th column of the channel matrix ($i.e.,$ $\bm{h}_{m}$) needs to be jointly performed over $n\in\{1,2,\ldots,N\}$.
Thus, the PDF of the extrinsic belief $\bm{l}^h_{k,m}$ for given $\bm{h}_{m}$ is given by

\begin{subequations}
\begin{align}
\label{eq:extrinsic_PDF_h_a}
p_{\bm{l}^h_{k,m}|\bm{h}_{m}}(\bm{l}^h_{k,m}|\bm{h}_{m}) &\! = \!\prod^N_{n=1}\prod^K_{i\neq n}p_{\tilde{y}_{m,ni}|h_{nm}}(\tilde{y}_{m,ni}|h_{nm}),\\
\label{eq:extrinsic_PDF_h_b}
&\hspace{-15ex} \propto e^{- \left(\bm{h}_{m}-\bm{\mu}^h_{k,m}\right)^{\rm H}{\bm{\Sigma}^h_{k,m}}^{\hspace{-2ex}-1}\left(\bm{h}_{m}-\bm{\mu}^h_{k,m}\right)},\\
&\hspace{-15ex} \propto \frac{1}{\pi^N|{\bm{\Sigma}^h_{k,m}}|}e^{- \left(\bm{h}_{m}-\bm{\mu}^h_{k,m}\right)^{\rm H}{\bm{\Sigma}^h_{k,m}}^{\hspace{-2ex}-1}\left(\bm{h}_{m}-\bm{\mu}^h_{k,m}\right)}.
\label{eq:extrinsic_PDF_h_c}
\end{align}
\end{subequations}

Notice that the expression in equation \eqref{eq:extrinsic_PDF_h_c} is in fact a complex multi-variate Gaussian distribution, that is
\begin{equation}
\label{eq:extrinsic_PDF_h}
p_{\bm{l}^h_{k,m}|\bm{h}_{m}}(\bm{l}^h_{k,m}|\bm{h}_{m}) \propto \hspace{-5ex}\underbrace{\mathcal{CN}_N\left(\bm{\mu}^h_{k,m},\bm{\Sigma}^h_{k,m}\right)}_\text{$N$-multivariate complex Gaussian distribution}\hspace{-5ex},
\end{equation}

where
\begin{subequations}
\begin{eqnarray}
&\bm{\mu}^h_{k,m}\triangleq \left[\hat{q}_{k,1m},\hat{q}_{k,2m},\ldots,\hat{q}_{k,Nm}\right]^{\rm T},&\\
&\bm{\Sigma}^h_{k,m}\triangleq \text{diag}\big(\psi^q_{k,1m},\psi^q_{k,2m},\ldots,\psi^q_{k,Nm}\big),&
\end{eqnarray}
with
\begin{eqnarray}
&\hat{q}_{k,nm} \triangleq \psi^q_{k,nm}\sum^K_{i\neq k}\frac{\hat{x}^{*}_{n,mi}\tilde{y}_{m,ni}}{v^h_{m,ni}},&\\
&\psi^q_{k,nm} \triangleq \bigg(\sum^K_{i\neq k} \frac{|\hat{x}_{n,mi}|^2}{v^h_{m,ni}}\bigg)^{-1}\hspace{-2ex}.&
\end{eqnarray}
\end{subequations}

Taking the expectation over the \acp{PDF} of the extrinsic beliefs given in equations \eqref{eq:extrinsic_PDF_x} and \eqref{eq:extrinsic_PDF_h}, respectively, soft estimates of $x_{mk}$ and $\bm{h}_m$ can  be respectively obtained as
\begin{subequations}
\begin{align}
&\!\!\hat{x}_{n,mk} = \sum_{x_q\in\mathscr{C}} \frac{x_q \cdot p_{l^x_{n,mk}|x_{mk}}(l^x_{n,mk}|x_{q})p_{x_{mk}}(x_q)}{\sum_{x^\prime_q\in\mathscr{C}}p_{l^x_{n,mk}|x_{mk}}(l^x_{n,mk}|x^\prime_q)p_{x_{mk}}(x^\prime_q)},\label{eq:expected_x_mean}\\
& \hat{\bm{h}}_{k,m} = \int_{\bm{h}_{m}} \bm{h}_{m} \frac{p_{\bm{l}^h_{k,m}|\bm{h}_{m}}(\bm{l}^h_{k,m}|\bm{h}_{m})p_{\bm{h}_{m}}(\bm{h}_{m})}{\int_{\bm{h}^\prime_{m}}p_{\bm{l}^h_{k,m}|\bm{h}^\prime_{m}}(\bm{l}^h_{k,m}|\bm{h}^\prime_{m})p_{\bm{h}_{m}}(\bm{h}^\prime_{m})},
\label{eq:expected_h_mean}
\end{align}
\end{subequations}
where the denominators are introduced to ensure that the integral of the posterior \acp{PDF} is equivalent to $1$.

Although a general closed-form expression of equation \eqref{eq:expected_x_mean} is not known for arbitrary discrete constellations, in the case of Gray-coded \ac{QPSK} it can be written as \cite{TakumiTC19}
\begin{equation}
\hat{x}_{n,mk} \!=\! \tfrac{1}{\sqrt{2}}\big(\!\tanh\!\big(\!\tfrac{\sqrt{2}\Re(\hat{r}_{n,mk})}{\psi^r_{n,mk}}\big) + j \tanh\big(\!\tfrac{\sqrt{2}\Im(\hat{r}_{n,mk})}{ \psi^r_{n,mk}}\!\big)\!\big),
\end{equation}
and its error variance estimate is given by
\begin{equation}\label{eq:residualnoise_var_x}
\psi^x_{n,mk} = 1 - |\hat{x}_{n,mk}|^2.
\end{equation}

In order to derive a closed-form expression for equation \eqref{eq:expected_h_mean}, we first simplify the effective \ac{PDF} such that
\begin{eqnarray}
P^h_{k,m}(\bm{h}_{m}) \triangleq  p_{\bm{l}^h_{k,m}|\bm{h}_{m}}(\bm{l}^h_{k,m}|\bm{h}_{m})p_{\bm{h}_{m}}(\bm{h}_{m}) = \frac{1}{\pi^N}\times &&\nonumber\\
&&\hspace{-52ex}\Bigg[ \tfrac{\lambda e^{-{\bm{\mu}^h_{k,m}}^{\!\!\!\!\!\!\mathrm{H}}\big(\!\bm{\Sigma}^h_{k\!\!\,,\!m}\! + \mathbf{\Gamma}_{\!m}\!\big)^{\!\!-\!1}{\!\!\!\!\bm{\mu}^h_{k,m}}}\mathcal{CN}_{\!N}\big(\!\mathbf{\Gamma}_{\!m}\! \big(\!\bm{\Sigma}^h_{k,m}+ \mathbf{\Gamma}_{\!m}\!\big)^{\!\!-\!1}\!\!\!\!\!,\mathbf{\Gamma}_{\!m}\!\big({\bm{\Sigma}^h_{k,m}} + \mathbf{\Gamma}_{\!m}\big)^{\!-\!1}{\!\!\bm{\Sigma}^h_{k,m}}\!\big)}{|\mathbf{\Gamma}_{\!m}+\bm{\Sigma}^h_{k,m}|}\nonumber\\
&&\hspace{-30ex} +  \tfrac{(1-\lambda)\delta(\bm{h}_{m})e^{- {\bm{\mu}^h_{k,m}}^{\!\!\!\!\!\!\mathrm{H}}{\bm{\Sigma}^h_{k,m}}^{\hspace{-2ex}-1}{\bm{\mu}^h_{k,m}}}}{|\bm{\Sigma}^h_{k,m} |}\Bigg].\label{eq:effective_PDF}
\end{eqnarray}

Also, the normalizing factor in the denominator of equation \eqref{eq:expected_h_mean} can be calculated by solving the integral over the entire $N$-dimensional complex field, which yields
\begin{align}
\label{eq:normalizing}
C^h_{k,m} & \triangleq\int_{\bm{h}^\prime_{m}}p_{\bm{l}^h_{k,m}|\bm{h}^\prime_{m}}(\bm{l}^h_{k,m}|\bm{h}^\prime_{m})p_{\bm{h}_{m}}(\bm{h}^\prime_{m}) \\
&=\frac{\lambda\exp\big( \!- {\bm{\mu}^h_{k,m}}^{\!\!\!\!\!\!\mathrm{H}}\left(\bm{\Sigma}^h_{k,m} + \mathbf{\Gamma}_m\right)^{-1}{\bm{\mu}^h_{k,m}}\big)}{\pi^N  |\mathbf{\Gamma}_m+\bm{\Sigma}^h_{k,m}|}\tau_{k,m},\nonumber
\end{align}
where 
\begin{subequations}
\label{eq:t_km}
\begin{equation}
\tau_{k,m} \triangleq 1 + \frac{1-\lambda}{\lambda} 
\exp\big(- \big(\pi^h_{k,m} - \psi^h_{k,m}\big)\big),\label{eq:ta}
\end{equation}
\vspace{-3ex}
\begin{equation}
\pi^h_{k,m} \triangleq {\bm{\mu}^h_{k,m}}^{\!\!\!\!\!\!\mathrm{H}}\big({\bm{\Sigma}^h_{k,m}}^{\hspace{-2ex}-1}  - \left(\bm{\Sigma}^h_{k,m} + \mathbf{\Gamma}_m\right)^{-1}\big){\bm{\mu}^h_{k,m}},
\label{eq:tb}
\vspace{-1ex}
\end{equation}
\begin{equation}
\psi^h_{k,m} \triangleq \log\left( \big|{\bm{\Sigma}^h_{k,m}}^{\hspace{-2ex}-1}\mathbf{\Gamma}_m+\mathbf{I}_{N} \big|\right).\label{eq:tc}
\end{equation}
\end{subequations}

For the sake of readability, we have moved the detailed derivations of equations \eqref{eq:effective_PDF} and \eqref{eq:normalizing} to Appendix \ref{app:effective_PDF}.

In light of the above, the soft replica for $\bm{h}_{nm}$ for a given effective distribution $p_{\bm{l}^h_{k,m}|\bm{h}_{m}}(\bm{l}^h_{k,m}|\bm{h}_{m})p_{\bm{h}_{m}}(\bm{h}_{m})$ can be obtained as
\begin{equation}
\hat{\bm{h}}_{k,m} = \frac{\mathbf{\Gamma}_m \big(\bm{\Sigma}^h_{k,m}+ \mathbf{\Gamma}_m\big)^{-1}}{\tau_{k,m}}{\bm{\mu}^h_{k,m}},
\end{equation}
and for the corresponding error variance, introducing $\bm{\Psi}^h_{k,m}\triangleq \diag{\psi^h_{k,1m},\psi^h_{k,2m},\ldots,\psi^h_{k,Nm}}$ allows us to write
\begin{eqnarray}
\label{eq:residualnoise_var_h}
\bm{\Psi}^h_{k,m} \hspace{-4.5ex}&&=\! \text{diag}\bigg(\!\int_{\bm{h}_{m}}\!\! \!\!\bm{h}_{m}\bm{h}^{\rm H}_{m}\frac{P^h_{k,m}(\bm{h}_{m})}{C_{k,m}}- \hat{\bm{h}}_{k,m}\hat{\bm{h}}^{\rm H}_{k,m}\bigg)\\
&& =\! (\!\tau_{k,m}\!-\!1\!)\text{diag}(\hat{\bm{h}}_{k,m}\hat{\bm{h}}^{\rm H}_{k,m})\!+\! \frac{\bm{\Sigma}^h_{k,m}\mathbf{\Gamma}_m({\bm{\Sigma}^h_{k,m}}\!\! +\! \mathbf{\Gamma}_m)^{\!-1}}{\tau_{k,m}}.\nonumber
\end{eqnarray}

\vspace{-3ex}
\subsection{Algorithm Description}
\label{sec:alg_descript}

In this subsection we summarize the belief propagation and consensus mechanisms described above and schematized in Figure \ref{fig:work_flow}, offering also detailed discussion on the algorithmic flow.
For starters, the schemes are concisely described in the fork of pseudo-codes in Algorithm \ref{alg:alg1} and Algorithm \ref{alg:alg2}, respectively.
For the sake of brevity, we define two sets of integers ($i.e.,$ $\mathscr{K}_p$ and $\mathscr{K}_d$), which are respectively given by
$\mathscr{K}_p\triangleq\{1,2\ldots,K_p\}$ and 
$\mathscr{K}_d\triangleq\{K_p+1,K_p+2\ldots,K\}$.

As shown in the pseudo-codes, Algorithm \ref{alg:alg1} requires five different inputs: the received signal matrix $\bm{Y}$, the pilot sequence $\bm{X}_p$, an initial guess of the channel matrix $\hat{\bm{H}}$, an initial guess of the estimation error variance $\hat{\bm{\Psi}}^h$ corresponding to $\hat{\bm{H}}$, and the maximum number of iterations $t_{\rm max}$; while Algorithm \ref{alg:alg2} is fed with the beliefs obtained from Algorithm \ref{alg:alg1}.
We point out that rough estimates $\hat{\bm{H}}$ and $\hat{\bm{\Psi}}^h$ can be obtained through state-of-the-art estimation algorithms proposed for grant-free systems \cite{ZhilinTWC19, LiangTSP18, CaireAsilomar2019}, although such mechanisms suffer from estimation inaccuracy in case of severely non-orthogonal pilot sequence ($K_p\ll M$), which is however desired from a system-level perspective in terms of time resource efficiency.  
In Section \ref{sec:initialization}, the initialization process adopted in this article will be discussed in details.

\begin{algorithm}[H]
\hrulefill
\begin{algorithmic}[1]
\vspace{-0.5ex}
\Statex {\bf{Input:}} $\bm{Y}$, $\bm{X}_p$, $\hat{\bm{H}}$, $\hat{\bm{\Psi}}^h$, $\lambda$, $t_{\rm max}$
\Statex {\bf{Output:}} $\forall k\in \mathscr{K}_d$, $\forall m$, $\forall n$: $\hat{x}_{n,mk}$, $\psi^x_{n,mk}$,
\Statex \hspace{8.5ex}$\forall k$, $\forall m$: $\hat{\bm{h}}_{k,m}$, $\bm{\Psi}^h_{k,m}$, $\forall m$: $\bm{\Gamma}_m$
\vspace{-1.5ex}
\Statex \hspace{-4ex}\hrulefill
\Statex  \fbox{\emph{Initialization}}
\vspace{0.5ex}
\State $\forall k\in \mathscr{K}_p$, $\forall m$, $\forall n$:
\Statex $\hat{x}_{n,mk}(1)=\big[\bm{X}_p\big]_{mk}$, $\psi^x_{n,mk}(1)=0$
%
\State $\forall k\in \mathscr{K}_d$, $\forall m$, $\forall n$: $\hat{x}_{n,mk}(1)=0$, $\psi^x_{n,mk}(1)=1$
%
\State $\forall k$, $\forall m$, $\forall n$:
\Statex $\hat{h}_{k,nm}(1)=\big[\hat{\bm{H}}\big]_{nm}$, $\psi^h_{k,nm}(1)=\big[\hat{\bm{\Psi}}^h\big]_{nm}$
\vspace{0.75ex}
\Statex \fbox{\emph{Main loop}}
\vspace{0.5ex}
\State {\bf for} $t=1,\ldots,t_{max}$
%
\State $\forall k$,\! $\forall m$,\! $\forall n$:\! $\tilde{y}_{m,nk}(t) \!=\! y_{nk} \!-\! \sum^M_{i\neq m}\!\hat{h}_{k,ni}(t)\hat{x}_{n,ik}(t)$
%
\State $\forall k$,\! $\forall m$,\! $\forall n$:\! $v^y_{m,nk}(t)\!=\!  \sum^{M}_{i\neq m} |\hat{h}_{k,ni}(t)|^2 \psi^x_{n,ik}(t)$
\NoNumber{\hspace{10ex}$ + (|\hat{x}_{n,ik}(t)|^2 \!+\! \psi^x_{n,ik}(t))\psi^h_{k,ni}(t) \!+\! N_0$}
\vspace{1ex}
\State $\forall k\in\mathscr{K}_d$,\! $\forall m$,\! $\forall n$:\! $v^x_{m,nk}(t) = v^y_{m,nk}(t)+\psi^h_{k,nm}(t)$\!
%
\State $\forall k$,\! $\forall m$,\! $\forall n$:\! $v^h_{m,nk}(t) \!=\! v^y_{m,nk}(t)\!+\!\gamma_{nm}\psi^x_{n,mk}(t)$
\vspace{0.5ex}
\State $\forall k\in\mathscr{K}_d$,\! $\forall m$,\! $\forall n$:
\NoNumber{\hspace{2ex}
$\psi^r_{n,mk}(t)=\big(\sum^N_{i\neq n} \frac{|\hat{h}_{k,im}(t)|^2}{v^x_{m,ik}(t)}\big)^{-1}$},
\NoNumber{\hspace{2.5ex}
$\hat{r}_{n,mk}(t)=\psi^r_{n,mk}(t)\sum^N_{i\neq n}\tfrac{\hat{h}^{*}_{k,im}(t)\tilde{y}_{m,ik}(t)}{v^x_{m,ik}(t)}$}
\vspace{0.5ex}
%
%
\State $\forall k$,\! $\forall m$,\! $\forall n$:
\NoNumber{\hspace{2ex}
$\psi^q_{k,nm}(t)=
\big(\sum_{i\neq k}^K \frac{|\hat{x}_{n,mi}(t)|^2}{v^h_{m,ni}(t)}\big)^{-1}$},
\NoNumber{\hspace{2.5ex}
$\hat{q}_{k,nm}(t)=\psi^q_{k,nm}(t)
\sum_{i\neq k}^K \frac{\hat{x}^{*}_{n,mi}(t)\tilde{y}_{m,ni}(t)}{v^h_{m,ni}(t)}$}
\vspace{0.5ex}
%
%
\State $\forall k$,\! $\forall m$:\! $\bm{\mu}^h_{k,m}(t) = [\hat{q}_{k,1m}(t),\ldots,\hat{q}_{k,Nm}(t)]^{\rm T}$
\vspace{0.25ex}
\State $\forall k$,\! $\forall m$:\! $\bm{\Sigma}^h_{k,m}(t) = \text{diag}(\psi^q_{k,1m}(t),\ldots,\psi^q_{k,Nm}(t))$
\vspace{1ex}
\State $\forall k$,\! $\forall m$:
\NoNumber{\hspace{2ex}$\pi^h_{k,m}(t) \!=\! {\bm{\mu}^h_{k,m}}^{\!\!\!\mathrm{H}}(t)\cdot$}
\NoNumber{\hspace{8ex}$({\bm{\Sigma}^h_{k,m}}^{\hspace{-2ex}-1}(t)  \!-\! (\bm{\Sigma}^h_{k,m}(t) \!+\! \mathbf{\Gamma}_m)^{-1}){\bm{\mu}^h_{k,m}}(t)$},
\vspace{0.25ex}
\NoNumber{\hspace{2ex}$\psi^h_{k,m}(t) \!=\! \log(|{\bm{\Sigma}^h_{k,m}}^{\hspace{-2ex}-1}(t)\mathbf{\Gamma}_m+\mathbf{I}_{N}|)$},
\vspace{0.25ex}
\NoNumber{\hspace{2.5ex}$\tau_{k,m}(t) \!=\! 1 \!+\! \frac{1-\lambda}{\lambda} 
\exp(-(\pi^h_{k,m}(t) - \psi^h_{k,m}(t)))$}
\vspace{0.75ex}
\State $\forall k$,\! $\forall m$:\! $\bar{\bm{h}}_{k,m}(t+1) \!=\! \frac{\mathbf{\Gamma}_m \big(\bm{\Sigma}^h_{k,m}(t)+ \mathbf{\Gamma}_m\big)^{-1}}{\tau_{k,m}(t)}{\bm{\mu}^h_{k,m}(t)}$
\vspace{0.5ex}
\State $\forall k$,\! $\forall m$:\! $\hat{\bm{h}}_{k,m}\!(t+1) \!=\! \eta\bar{\bm{h}}_{k,m}(t\!+\!1) \!+\! (1\!-\!\eta) \hat{\bm{h}}_{k,m}(t)$
\vspace{0.5ex}
\State $\forall k$,\! $\forall m$:\! $\bar{\bm{\Psi}}^h_{k,m}(t+1)=(\tau_{k,m}(t)-1)\cdot$ 
\NoNumber{$\text{diag}(\bar{\bm{h}}_{k,m}(t)\bar{\bm{h}}^{\rm H}_{k,m}(t)) + \frac{\bm{\Sigma}^h_{k,m}(t)\mathbf{\Gamma}_m({\bm{\Sigma}^h_{k,m}}(t) + \mathbf{\Gamma}_m)^{-1}}{\tau_{k,m}(t)}$}
\vspace{0.5ex}
\State $\forall k$,\! $\forall m$:\! $\bm{\Psi}^h_{k,m}\!(t\!+\!1) \!=\! \eta\bar{\bm{\Psi}}^h_{k,m}\!(t\!+\!1) \!+\! (1\!-\!\eta) \bm{\Psi}^h_{k,m}\!(t)$
\vspace{0.75ex}
\State $\forall k\in\mathscr{K}_d$,\! $\forall m$,\! $\forall n$:\! $\bar{x}_{n,mk}(t+1)=\tfrac{\tau^{-1}_{k,m}(t)}{\sqrt{2}}\cdot$ \NoNumber{\footnotesize\hspace{-1.2ex}$\big(\!\tanh\!\big(\!\tfrac{\sqrt{2}\gamma(t)\Re(\hat{r}_{n,mk}(t))}{\psi^r_{n,mk}(t)}\big) \!+ j\!\cdot\! \tanh\!\big(\!\tfrac{\sqrt{2}\gamma(t)\Im(\hat{r}_{n,mk}(t))}{\psi^r_{n,mk}(t)}\big)\!\big)$}
\vspace{0.5ex}
\State $\forall k\in\mathscr{K}_d$,\! $\forall m$,\! $\forall n$:\! 
\NoNumber{\hspace{3ex}$\hat{x}_{n,mk}(t\!+\!1) \!=\!\eta\bar{x}_{n,mk}(t\!+\!1) \!+\! (1\!-\!\eta) \hat{x}_{n,mk}(t)$}
\vspace{0.5ex}
\State $\forall k\in\mathscr{K}_d$,\! $\forall m$,\! $\forall n$: $\psi^x_{n,mk}(t+1)$ 
\NoNumber{\hspace{3ex}$={\eta\cdot \tau^{-1}_{k,m}(1 - |\bar{x}_{n,mk}(t)|^2)} + (1-\eta)\psi^x_{n,mk}(t)$}
\vspace{0.5ex}
\State {\bf end for}
\caption[]{:\\ Bilinear \ac{GaBP} (Part1: Belief Consensus)}
\label{alg:alg1}
\end{algorithmic}
\end{algorithm}
\setlength{\textfloatsep}{5pt}

In turn, the outputs of Algorithm \ref{alg:alg2} are the following three quantities: an estimated symbol matrix $\hat{\bm{X}}$, an estimated channel matrix $\hat{\bm{H}}$, and estimates of active-user indexes $\hat{\mathscr{A}}$.
As for $\hat{\bm{X}}$ and $\hat{\bm{H}}$, they are simply obtained by combining all the beliefs ($i.e.,$ consensus), while $\hat{\mathscr{A}}$ is determined by following a certain activity detection policy based on the estimated channel $\hat{\bm{H}}$.
The activity detection policy considered in this article will be described in Section \ref{sec:ADP} in further details.

Although most of the algorithmic flow in Algorithm \ref{alg:alg1} and \ref{alg:alg2} follows the belief exchange policy described in Section \ref{sec:FN} and \ref{sec:VN} above, we have brought two belief manipulation techniques ($i.e.,$ damping \cite{RajanMIMO14} and scaling\cite{TakumiTC19}) to Algorithm \ref{alg:alg1} in order to further improve the estimation accuracy and escape from local optima.

This is due to the fact that when the Gaussian approximation assumed in equation \eqref{eq:Gaussian_belief} does not sufficiently describe the actual stochastic behavior of the effective noise, the accuracy of soft-replicas is degraded by resultant belief outliers caused by the aforementioned approximation gap, leading to unignorable estimation performance deterioration.

This error propagation effect becomes predominant when the pilot length decreases, which is exactly the scenario the article aims at.

\begin{algorithm}[H]
\hrulefill
\begin{algorithmic}[1]
\vspace{-0.7ex}
\Statex {\bf{Input:}} $\bm{Y}$, $\forall k$, $\forall m$, $\forall n$: $\hat{x}_{n,mk}$, $\psi^x_{n,mk}$,
\Statex \hspace{10.3ex}$\forall k$, $\forall m$: $\hat{\bm{h}}_{k,m}$, $\bm{\Psi}^h_{k,m}$, $\forall m$: $\bm{\Gamma}_m$

\Statex {\bf{Output:}} $\hat{\bm{X}}$, $\hat{\bm{H}}$, $\hat{\mathscr{A}}$
\vspace{-1.7ex}
%
\Statex \hspace{-4ex}\hrulefill
\State $\forall k$,\! $\forall m$,\! $\forall n$:\! $\tilde{y}_{m,nk} = y_{nk} \!-\! \sum^M_{i\neq m}\!\hat{h}_{k,ni}\hat{x}_{n,ik}$
\vspace{0.5ex}
\State $\forall k$,\! $\forall m$,\! $\forall n$:\! $v^y_{m,nk}= $
\NoNumber{\!${\mathop {\sum} \limits_{i\neq m}} \!|\hat{h}_{k,ni}|^2 \psi^x_{n,ik} \!+\! (|\hat{x}_{n,ik}|^2 \!+ \psi^x_{n,ik})\psi^h_{k,ni} \!+\!\! N_0$}
\vspace{1ex}
\State $\forall k\in\mathscr{K}_d$,\! $\forall m$,\! $\forall n$:\! $v^x_{m,nk} = v^y_{m,nk}+\psi^h_{k,nm}$\!
\vspace{0.5ex}
\State $\forall k$,\! $\forall m$,\! $\forall n$:\! $v^h_{m,nk} = v^y_{m,nk}\!+\!\gamma_{nm}\psi^x_{n,mk}$
\vspace{1ex}
\State $\forall k\in\mathscr{K}_d$,\! $\forall m$:\! $\psi^r_{mk} = \Big(\sum^N_{i=1}\! \frac{|\hat{h}_{k,im}|^2}{v^x_{m,ik}}\Big)^{-1}$
\vspace{0.5ex}
\NoNumber{\hspace{14ex}$\hat{r}_{mk} = \psi^r_{mk}\sum^N_{i=1}\frac{\hat{h}^{*}_{k,im}\tilde{y}_{m,ik}}{v^x_{m,ik}}$}
\vspace{0.5ex}
\State $\forall m$,\! $\forall n$: $\psi^q_{nm}= \Big(\sum^K_{i=1} \tfrac{|\hat{x}_{n,mi}|^2}{v^h_{m,ni}}\Big)^{-1}$
\vspace{0.5ex}
\NoNumber{\hspace{8.6ex}$\hat{q}_{nm}=\psi^q_{nm}\cdot\sum^K_{i=1}\tfrac{\hat{x}^{*}_{n,mi}\tilde{y}_{m,ni}}{v^h_{m,ni}}$}
\vspace{1ex}
\State $\forall k\in\mathscr{K}_d$,\! $\forall m$:\!  \NoNumber{$\bar{x}_{mk}\!=\!\tfrac{1}{\sqrt{2}}\big(\!\tanh\!\big(\!\tfrac{\sqrt{2}\Re(\hat{r}_{mk})}{\psi^r_{mk}}\big) \!+ j\!\cdot\! \tanh\!\big(\!\tfrac{\sqrt{2}\Im(\hat{r}_{mk})}{\psi^r_{mk}}\big)\!\big)$}
\vspace{1ex}
\State $\forall k\in\mathscr{K}_d$,\! $\forall m$: $\hat{x}_{mk} = {\mathop {\mathrm{argmin}} \limits_{x_q\in\mathscr{C}}}\:\: |x_q-\bar{x}_{mk}|$ 
\vspace{1ex}
\State $\forall m$:\! $\bm{\mu}^h_{m} = [\hat{q}_{1m},\ldots,\hat{q}_{Nm}]^{\rm T}$
\vspace{0.25ex}
\State $\forall m$:\! $\bm{\Sigma}^h_{m} = \text{diag}(\psi^q_{1m},\ldots,\psi^q_{Nm})$
\vspace{1ex}
\State $\forall m$:\! $\pi^h_{m} = {\bm{\mu}^h_{m}}^{\mathrm{H}}({\bm{\Sigma}^h_{m}}^{-1} - (\bm{\Sigma}^h_{m} + \mathbf{\Gamma}_m)^{-1}){\bm{\mu}^h_{m}}$
\vspace{0.5ex}
\NoNumber{\hspace{4ex}$\psi^h_{m} = \log(|{\bm{\Sigma}^h_{m}}^{-1}\mathbf{\Gamma}_m+\mathbf{I}_{N}|)$},
\vspace{0.5ex}
\NoNumber{\hspace{4.5ex}$\tau_{m} = 1 + \frac{1-\lambda}{\lambda} 
\exp(-(\pi^h_{m} - \psi^h_{m}))$}
\vspace{1ex}
\State $\forall m$:\! $\hat{\bm{h}}_{m} = \frac{\mathbf{\Gamma}_m \big(\bm{\Sigma}^h_{m}+ \mathbf{\Gamma}_m\big)^{-1}}{\tau_{m}}{\bm{\mu}^h_{m}}$
\vspace{1ex}
\State $\hat{\mathscr{A}} = \text{ActivityDetectionPolicy}(\hat{h}_{nm})$
\vspace{1ex}
\caption[]{:\\ Bilinear \ac{GaBP} (Part2: Hard Decision)}
\label{alg:alg2}
\end{algorithmic}
\end{algorithm}
\setlength{\intextsep}{3pt}
\vspace{2ex}

Following \cite{ParkerTSP14, ItoGC20}, we have applied damping to line 15, 17 and 19--20 of Algorithm \ref{alg:alg1} with the damping factor $\eta\in[0,1]$, which tends to prevent the algorithm from converging to local minima by forcing a slow update of soft-replicas, whereas belief scaling is adopted in line 18 of Algorithm \ref{alg:alg1} with parameter $\gamma(t)$, which in turn adjust the reliability of beliefs ($i.e.,$ harnessing harmful outliers).
The dynamics is designed to be a linear function of the number of iterations, that is,
\begin{equation}
\label{eq:ASB_gamma}
\gamma(t) = \frac{t}{t_{\rm max}}.
\end{equation}

Besides the above, as shown in Section \ref{sec:VN}, soft estimates of $x_{mk}$ are obtained without considering the user activity ($i.e.,$ row-sparsity) by imposing user activity detection upon the channel estimation process as described in equation \eqref{eq:expected_h_mean}, indicating that such row-sparsity of $\bm{X}$ needs to be incorporated so as to avoid inconsistency with column-sparsity of $\bm{H}$.

Ironing out this issue, we leverage the sparsity factor $\tau_{k,m}$ given in equation \eqref{eq:t_km} in line 18 and 20 of Algorithm \ref{alg:alg1}, which tends to be $1$ when active and to be $\infty$ otherwise, such that rows of $\bm{X}$ corresponding to non-active columns of $\bm{H}$ are suppressed, maintaining the consistency.

%
%

\subsection{Initialization}
\label{sec:initialization}

Due to the fact that bilinear inference problems are sensitively affected by the initial values of the solution variables, a reasonable initialization method is required so that the algorithm accurately estimates the channel, the informative data and the activity pattern simultaneously.
However, one may also notice that due to the severe non-orthogonality of the pilot sequence ($i.e.,$ $K_p \ll |\mathscr{A}| \ll M$) for overhead reduction, the accuracy of such an initial guess is not reliable enough.  

In light of the above, although several approaches developed for grant-free access such as covariance-based methods \cite{ZhilinICC19,CaireAsilomar2019} can be considered to produce initial $\hat{\bm{H}}$ and $\hat{\bm{\Psi}}^h$, in this article we have leveraged \ac{MMV-AMP} \cite{LiangTSP18} from a computational complexity perspective\footnote{Please refer to \cite{AlexanderArxiv19} for complexity analyses between existing grant free schemes for further details.}, which can be simply applied to equation \eqref{eq:RXsignals} by regarding the first $K_p$ columns of $\bm{Y}$ and the pilot matrix as the effective received signal matrix and its measurement matrix, respectively.

\subsection{Activity Detection Policy}
\label{sec:ADP}

In this section, we describe how to identify user activity patterns based on the belief consensus performed in Algorithm \ref{alg:alg2}.
Due to the fact that miss-detection and false alarms are in a trade-off relationship as shown in the grant-free literature, such an activity detection policy is affected by system and user requirements, indicating that one needs to adopt a suitable criterion depending on the situation in practical implementations.

With that in mind, we simply consider the log likelihood ratio method based on estimated channel quantities, which thanks to uncorrelated Gaussianity of the channel and residual estimation error, can be written as
\begin{equation}\label{eq:LLR}
\text{LLR}_m \triangleq \ln \frac{\prod^N_{n=1}\mathcal{CN}\left(0, \gamma_{nm}+\psi^h_{nm}|\hat{h}_{nm}\right)}{\prod^N_{n=1}\mathcal{CN}\left(0, \psi^h_{nm}|\hat{h}_{nm}\right)},
\end{equation}
where $\text{LLR}_m$ is the log likelihood ratio corresponding to the $m$-th column and $\prod^N_{n=1}$ is introduced to perform consensus over the receive antenna dimension.

After some basic manipulations, the log likelihood criterion can be simplified to  
\begin{equation}\label{eq:LLR_quad}
\text{LLR}_m = p_{\rm active} - p_{\rm inactive}  \mathop{\gtrless}_{\rm inactive}^{\rm active} 0,
\end{equation}
where 
\begin{subequations}
\begin{eqnarray}
&p_{\rm active} \triangleq \sum^N_{n=1} - \frac{|\hat{h}_{nm}|^2}{\gamma_{nm}+\psi^h_{nm}} + \ln \left(\tfrac{1}{\pi(\gamma_{nm}+\psi^h_{nm})}\right),&\nonumber\\
&p_{\rm inactive} \triangleq \sum^N_{n=1} - \frac{|\hat{h}_{nm}|^2}{\psi^h_{nm}} + \ln \left(\tfrac{1}{\pi\psi^h_{nm}}\right).&\nonumber
\end{eqnarray}
\end{subequations}

\section{Simulation Results}
\label{sec:simulation}

In this section, we evaluate via software simulation the proposed method in terms of \ac{BER}, effective throughput, \ac{NMSE}, and \ac{AUD} performance. 

\subsection{Simulation Setup}
\label{sec:setup}

Throughout this performance assessment section, we consider the following simulation setup unless otherwise specified.
The number of receive antennas is set to $N=100$, which are distributed over a square of side of 1000 [m] in a square mesh fashion, where $M=100$ potential users are accommodated in each subcarrier.
%
%
It is assumed that $50$\% of the $M$ users become active during each \ac{OFDM} frame, $i.e.,$ $|\mathscr{A}|=M/2=50$, while $K$ and $K_p$ are considered to be $K\in\{140,280\}$ and $K_p=14$ depending on the employed subcarrier spacing\footnote{A scenario with $K=140$ and $K=280$ corresponds to \ac{OFDM} subcarrier spacing of $15$ [kHz] and $30$ [kHz], respectively. As shown in Figure \ref{fig:Model}, $K_p=14$ indicates that only one \ac{OFDM} slot is utilized as pilot and the rest as data transmission, indicating that this situation imposes the most severe scenario as the pilot length is minimum.}.
It is further worth-noting that since we accommodate $M=100$ users with overhead length $K_p=14$, a significant amount of overhead reduction can be achieved.
Although one might concern about performance degradation due to the resultant severe non-orthogonality, we dispel such concerns throughout this section by demonstrating that the bilinear inference method employed here is able to handle the non-orthogonality. 

The transmit power range at each uplink user is determined based on the experimental assessments studied in \cite{ParamanandaAccess17}, where the transmit power of each uplink user is limited by $16$ [dBm].
Furthermore, Gray-coded \ac{QPSK} modulation is assumed to be employed at each user, whereas the noise floor $N_0$ at each \ac{AP} is assumed to be modeled as 
\begin{equation}
\sigma^2_u = 10\log_{10}\left(1000 \kappa T \right) + \text{NF} + 10\log_{10}\left(W\right) \text{[dBm]},
\end{equation}
where $\kappa$ is the Boltzmann's constant, $T = 293.15$ denotes the physical temperature at each \ac{AP} in kelvins, the noise figure \text{NF} is assumed to be 5 [dB] and $W$ expresses the subcarrier bandwidth.

As described in Section \ref{sec:pilot}, the pilot structure is designed via quadratic \ac{CSIDCO} in order to mitigate pilot contamination effects as much as possible even in case of severely non-orthogonal scenarios such as one considered in this section.  
Regarding the effective throughput performance, we adopt the definition proposed in \cite[Def. 1]{DongxuInfocom03}, which is given by
\begin{equation}\label{eq:eff_throughput}
    \text{R}_{\rm eff} \triangleq (1-P_e)\cdot K_d \cdot b 
\end{equation}
where $P_e$ denotes the block (packet) error rate and $b$ is the number of bits per \ac{OFDM} symbol.

Regarding algorithmic parameters, the maximum number of iterations in Algorithm \ref{alg:alg1} is set to $t_{\rm max}=32$ and the damping factor $\eta$ is $0.5$, while the belief scaling factor $\gamma(t)$ is defined in equation \eqref{eq:ASB_gamma}.

\subsection[]{\ac{MUD}}
\label{sec:Sim_MUD}

The \ac{MUD} performance of the proposed algorithm is studied in terms of uncoded \ac{BER} as a function of transmit power.
In order to take into account \ac{MD} effects on data detection, we count not only bits received in error but also the number of lost bits due to missing user activity, that is,
\begin{equation}
    \text{BER} = \frac{P^1_e + P^2_e}{\text{Total number of bits}}
\end{equation}
where $P^1_e$ denotes the number of errors due to failure of symbol detection and $P^2_e$ is the number of bits that have been lost due to failure of user detection.

Since there are no existing cell-free scheme with grant-free access which do not rely on spreading data sequences as described in equation \eqref{eq:RXsignals}, we consider the \ac{MMV-AMP} algorithm as a state-of-the-art method to carry out \ac{JACE} with basis of non-orthogonal pilot sequences $\bm{X}_p$, remarking that this receiver is widely employed in related literature \cite{YangWCL18, HaraAccess19, ShuchaoTWC2020, WeijieTCom20}.
For the same reason (of lack of a direct equivalent competitor), we also compare the performance of our method against an idealized system in which perfect \ac{CE} and \ac{AUD} is assumed, with signal detection performed by the \ac{GaBP} algorithm.

\begin{figure}[H]
    \centering
    \begin{subfigure}[b]{0.48\textwidth}
        \includegraphics[width=\textwidth]{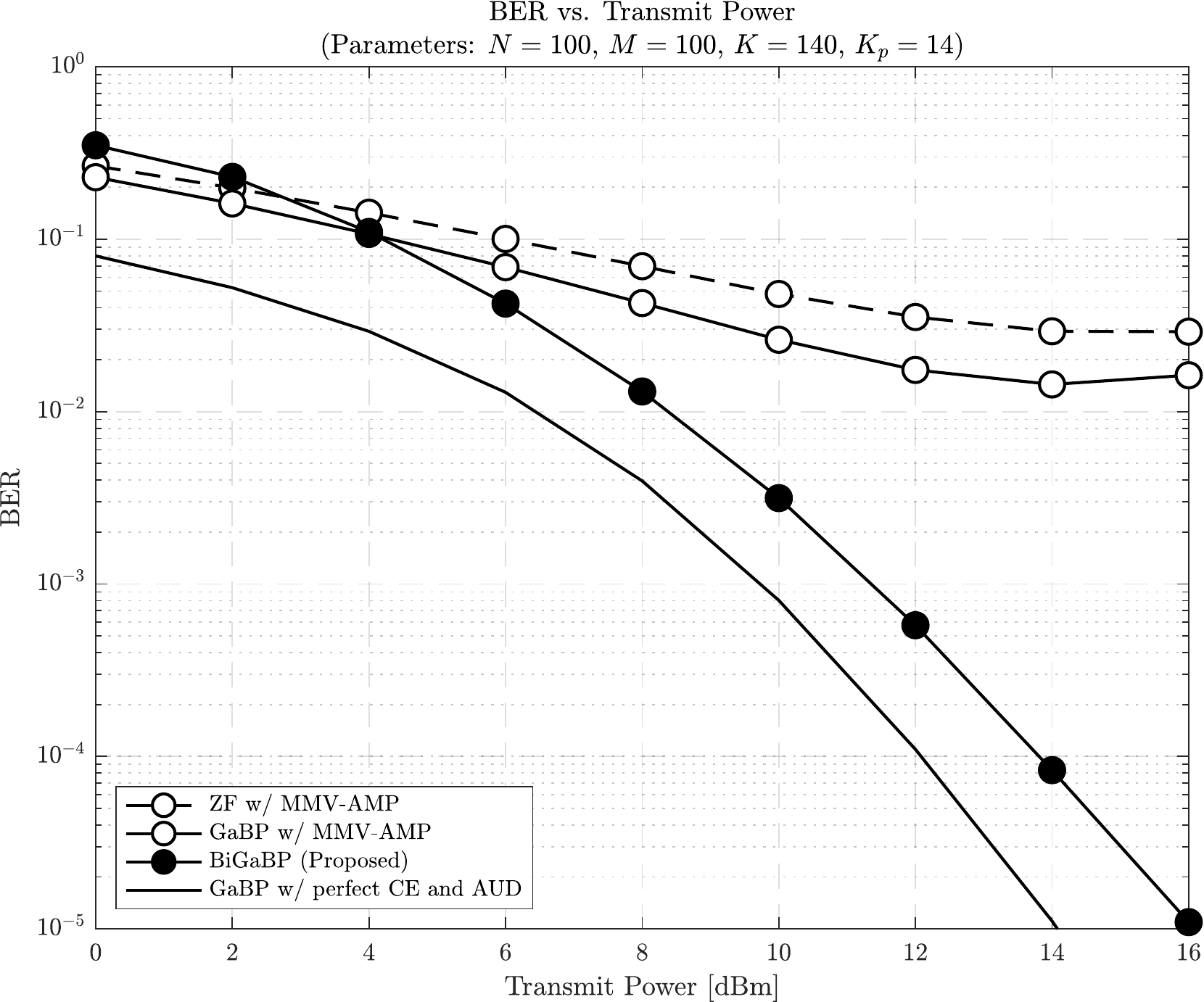}
        \caption{$15$ kHz subcarrier spacing ($N=M=100$ and $K_p=14$)}
        \label{fig:ber_15kHZ}
    \end{subfigure}
        ~ 
        \begin{subfigure}[b]{0.48\textwidth}
        \includegraphics[width=\textwidth]{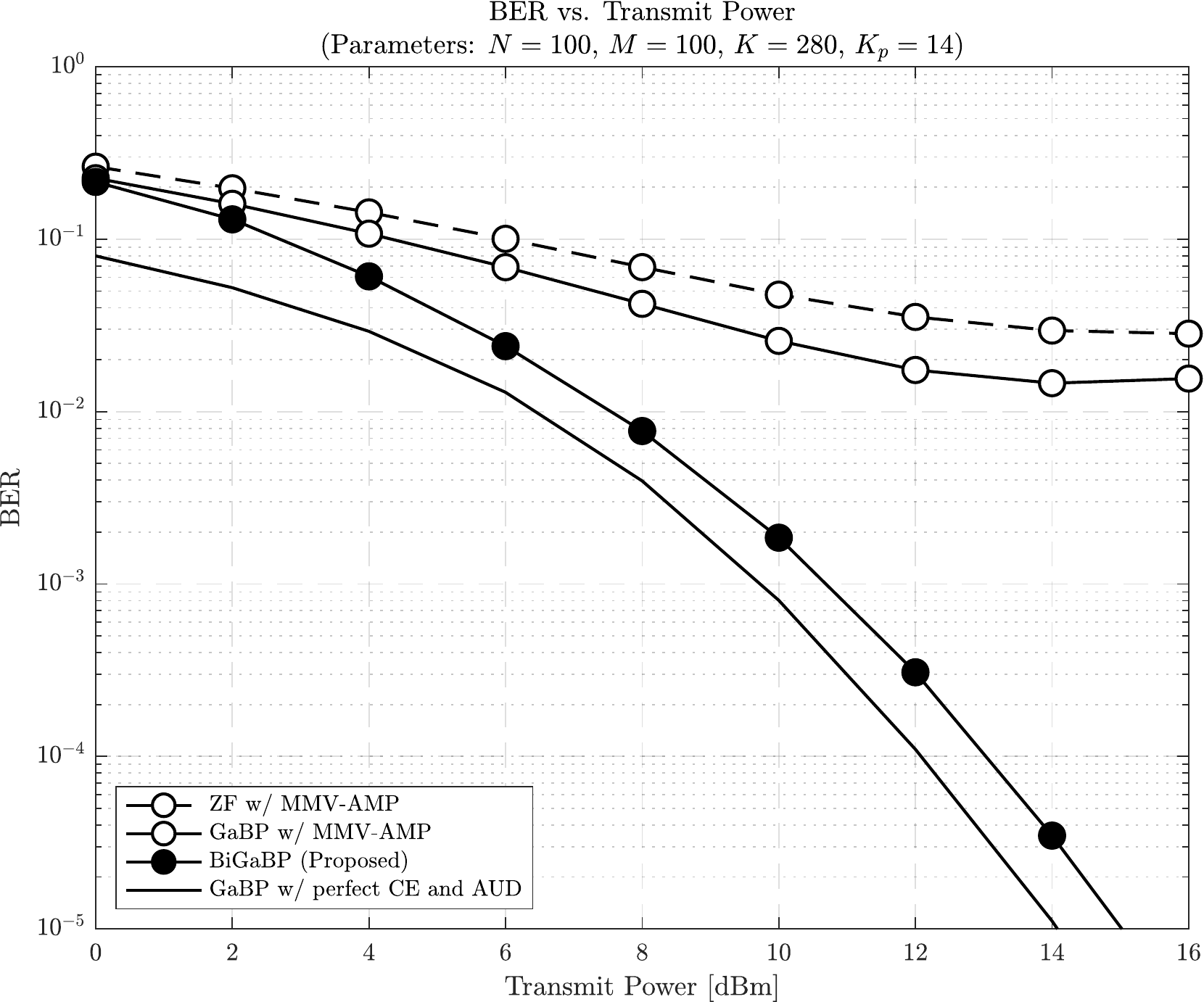}
        \caption{$30$ kHz subcarrier spacing ($N=M=100$ and $K_p=14$)}
        \label{fig:ber_30kHZ}
            \vspace{1ex}
    \end{subfigure}
    \caption{BER comparisons as a function of transmit power}
    \label{fig:ber}
\end{figure}

Our assessment starts with Figure \ref{fig:ber}, where the \ac{BER} performance of the proposed method is compared not only to the state-of-the-art but also to the ideal performance, for different subcarrier spacing scenarios ($i.e.,$ $K\in\{140,280\}$).

The state-of-the-art methods compared are the linear \ac{ZF} \ac{MIMO} detector and the \ac{GaBP} message passing \ac{MIMO} detector, followed by \ac{MMV-AMP}-based \ac{CE} with the aid of the non-orthogonal pilot sequence.
As for the ideal performance, the \ac{GaBP} \ac{MIMO} detector with perfect knowledge of \ac{CE} and \ac{AUD} at the receiver is employed, which can be thus considered extreme lower bound to the proposed method.

As can be seen from both figures, state-of-the-art methods suffer from high error floors, which stem from the poor \ac{CE} and \ac{AUD} performances by \ac{MMV-AMP}, caused by the severely overloaded condition.
In fact the aspect ratio (number of users over number of pilot symbols) of the pilot matrix is $\frac{M}{K_p}=\frac{100}{14}\approx 7.1428$, indicating a highly non-orthogonal condition.
Although only $m=50$ users out of $M=100$ are assumed to be active in each coherent frame, the overloading ratio is still sufficiently high to hinder \ac{CE} and \ac{AUD}.

\vspace{-10ex}

In contrast, the proposed method enjoys a water-falling curve in terms of \ac{BER} for the both situations, which can be achieved by taking advantage of the pseudo-orthogonality of the data structure.
This advantage can be confirmed from the fact that increasing the total symbol length from $K=140$ to $K=280$ while fixing the pilot length to be $K_p=14$ can indeed enhance the detection performance as shown in Figure \ref{fig:ber_15kHZ} and \ref{fig:ber_30kHZ}.
One may readily notice from the above that the corresponding \ac{CE} performance can be also improved due to the same logic, which is offered in Section \ref{sec:Sim_CE} below.

\subsection{Effective Throughput}
\label{sec:Sim_Throughput}

In light of the definition given in equation \eqref{eq:eff_throughput}, we next investigate the effective throughput performance per each \ac{OFDM} frame of the proposed method.
Please note that since the \ac{OFDM} frame corresponds to $10$ [ms], the effective throughput implies the number of successfully delivered bits within a resource block of $10$ [ms] times an \ac{OFDM} subcarrier.

\begin{figure}[H]
        \centering
    \begin{subfigure}[b]{0.48\textwidth}
        \includegraphics[width=\textwidth]{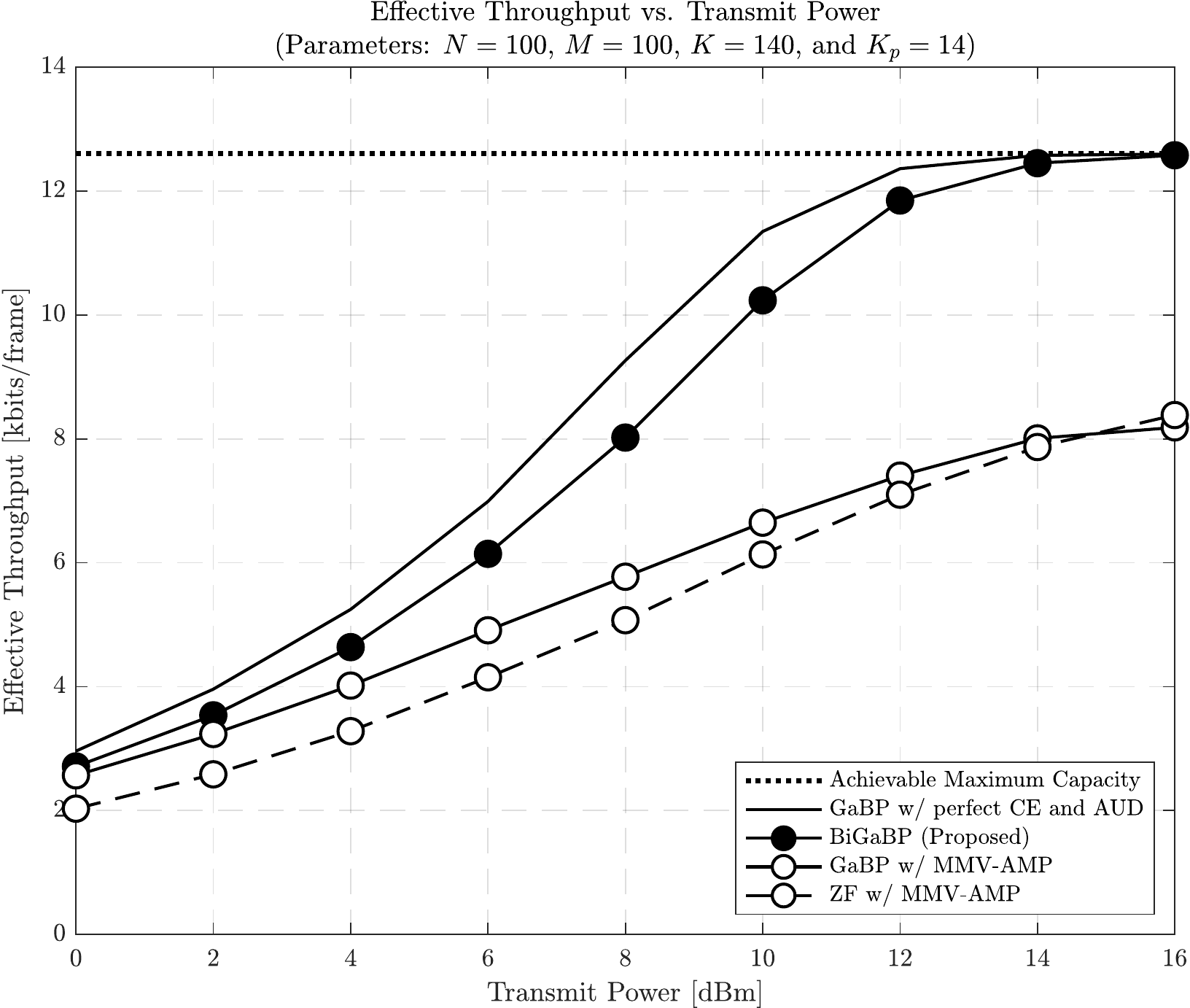}
        \caption{$15$ kHz subcarrier spacing ($N=M=100$ and $K_p=14$)}
    \end{subfigure}
        ~ 
      \begin{subfigure}[b]{0.48\textwidth}
        \includegraphics[width=\textwidth]{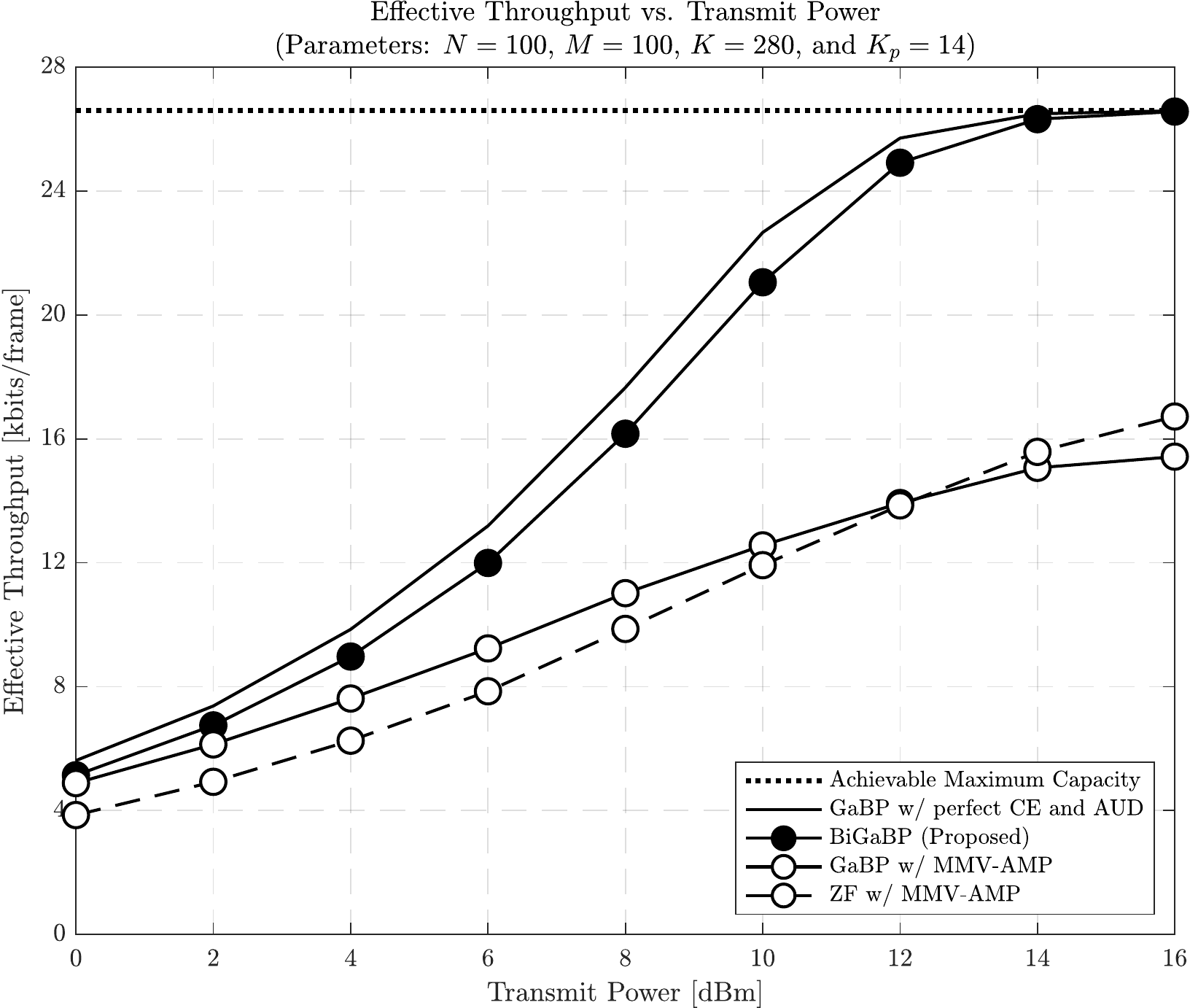}
     \caption{$30$ kHz subcarrier spacing ($N=M=100$ and $K_p=14$)}
    \end{subfigure}
    \caption{Effective throughput comparisons as a function of transmit power}
    \label{fig:rate}
\end{figure}

Simulation results showing the effective throughput achieved with the proposed scheme and compared alternatives are offered in Figure \ref{fig:rate} for different subcarrier spacing setups.
Note that the unit of the vertical axis is set to \emph{kilo}bits per frame for the sake of readability.
Furthermore, we also implicitly measure the packet (block) error performance of the methods as shown in equation \eqref{eq:eff_throughput}, which is often used for practical performance assessment.
Finally, in addition to the three counterparts considered in the previous section, we also offer in both figures the system-level achievable maximum data rate as reference, which is determined by 
\vspace{-1ex}
\begin{equation}
    \text{Maximum Capacity} \triangleq K_d\cdot |\mathscr{A}| \cdot |\mathscr{C}|\:\:\:\: \text{[bits/frame]},
\vspace{-1ex}
\end{equation}
where $|\mathscr{C}|$ denotes the number of bits per symbol.

As expected from the discussion of the previous section, it is found that the two state-of-the-art alternatives are incapable of successfully delivering bits transmitted by the active users even with sufficiently high transmit power.

In contrast, the proposed method dynamically follows the same improvement in performance with transmit power as the idealized receiver, approaching the achievable capacity as the transmit power increases.
It is furthermore seen that, as inferred in Section \ref{sec:Sim_MUD}, the throughput gap from the idealized scheme narrows as the subcarrier spacing increases, which is again due to the pseudo-orthogonality of the data sequence.

\vspace{-1ex}
\subsection[]{\ac{CE}}
\label{sec:Sim_CE}
\vspace{-0.5ex}

In addition to the above, the \ac{CE} performance of the proposed method is assessed in this section as a function of transmit power for different data lengths, so that one may observe that the performance improvement described in the above sections is, at least in part, induced by the resultant \ac{CE}.

\begin{figure}[H]
        \centering
    \begin{subfigure}[b]{0.48\textwidth}
        \includegraphics[width=\textwidth]{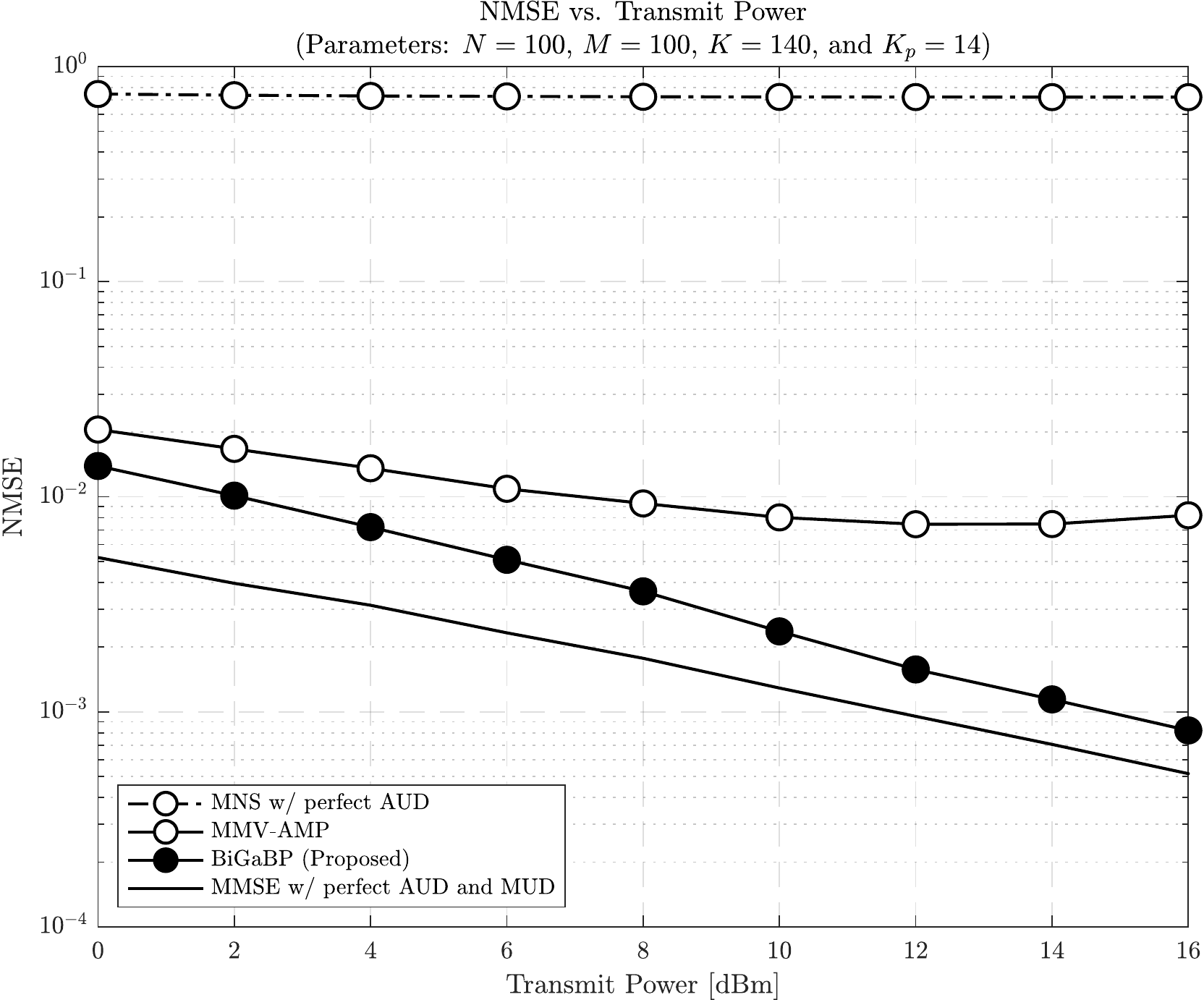}\vspace{-1ex}
        \caption{$15$ kHz subcarrier spacing ($N=M=100$ and $K_p=14$)}
        \label{fig:nmse_15kHZ}
    \end{subfigure}
        ~ 
        \begin{subfigure}[b]{0.48\textwidth}
        \includegraphics[width=\textwidth]{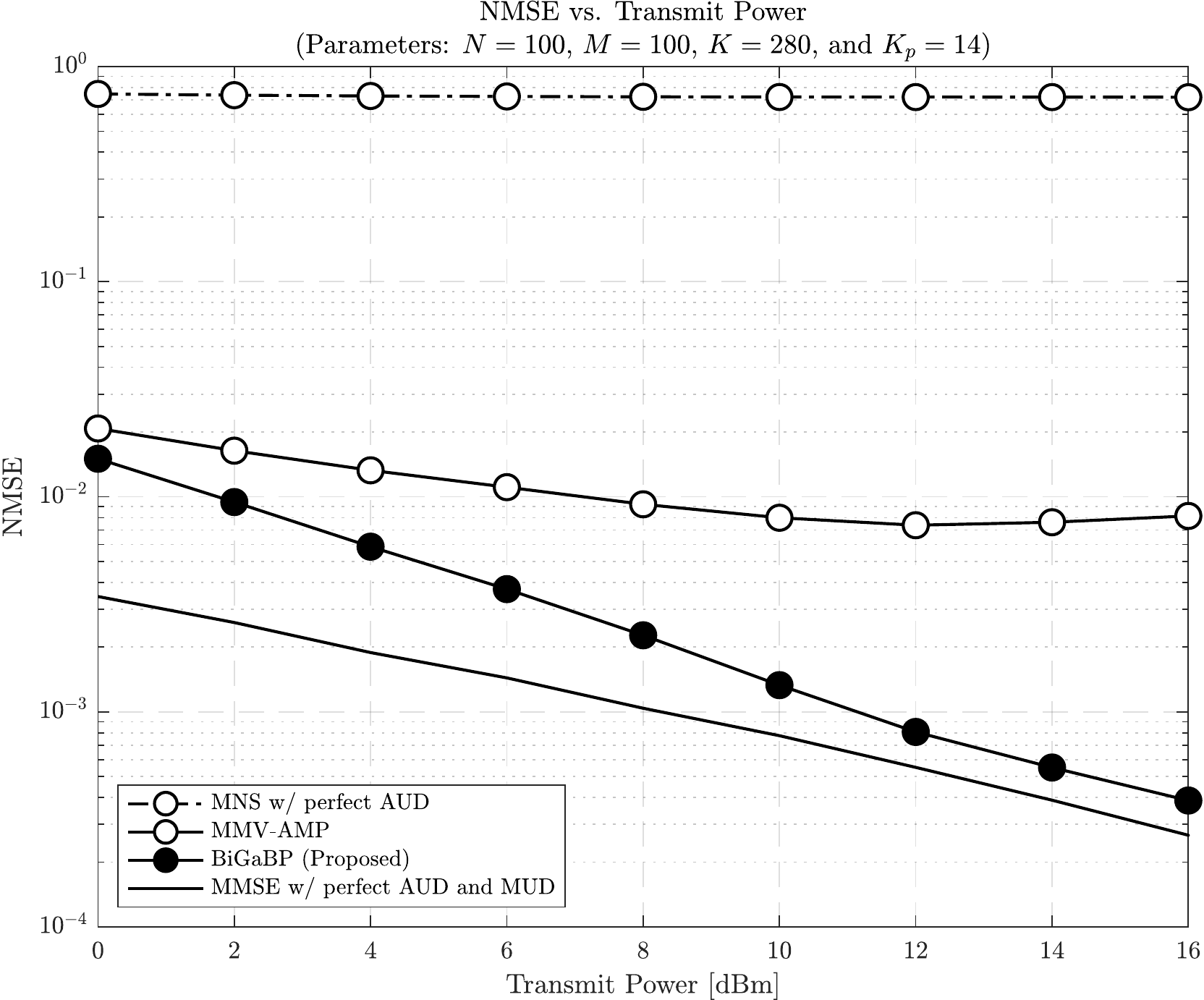}\vspace{-1ex}
     \caption{$30$ kHz subcarrier spacing ($N=M=100$ and $K_p=14$)}
     \label{fig:nmse_30kHZ}
     \vspace{0.5ex}
    \end{subfigure}
    \vspace{-1ex}
    \caption[]{\Ac{NMSE} comparisons as a function of transmit power}
    \label{fig:nmse}
\end{figure}

To this end, the \ac{NMSE} performance of the proposed method for $K=140$ and $K=280$ is offered in Figure \ref{fig:nmse_15kHZ} and \ref{fig:nmse_30kHZ}, respectively, where the \ac{NMSE} is defined as
\vspace{-0.5ex}
\begin{equation}
  \text{NMSE} \triangleq \frac{\|\bm{H} - \hat{\bm{H}}\|^2_{\rm F}}{\|\bm{H}\|^2_{\rm F}},
  \vspace{-0.5ex}
\end{equation}
assuming a fixed pilot length $K_p=14$.

As for methods to compare, we have adopted not only \ac{MMV-AMP} but also \ac{MNS} that is known to be a method to seek a closed-form unique \ac{CE} solution in case of a non-orthogonal pilot sequence \cite{ItoGC20}, while employing the \ac{MMSE} performance with perfect knowledge of \ac{AUD} and \ac{MUD} at the receiver as reference.
Please note that since the non-Bayesian approach, which takes advantage of the sample covariance of the received signals in order to detect user activity patterns, aims at only \ac{AUD}, the resultant performance in terms of \ac{CE} can be lower-bounded by \ac{MNS} with perfect \ac{AUD}.

With that in mind, it can be observed from Figure \ref{fig:nmse_15kHZ} and \ref{fig:nmse_30kHZ} that the proposed method can indeed improve the \ac{CE} performance and approach the unachievable \ac{MMSE} performance with perfect \ac{AUD} and \ac{MUD}, maintaining a similar gradient with that of the \ac{MMSE}, whereas  \ac{MMV-AMP} and \ac{MNS} suffer from a relatively high error floor due to the non-orthogonality of the pilot, although \ac{MMV-AMP} appears to offer moderate performance in comparison with \ac{MNS}.  

Thanks to the pseudo-orthogonality of the data structure, the proposed method with $30$ kHZ subcarrier spacing again outperforms its own \ac{NMSE} with $15$ kHZ subcarrier spacing.
Furthermore, it can be mentioned that due to the sufficiently high \ac{CE} accuracy of the proposed method ($i.e.,$ $\text{NMSE}\in[10^{-3},10^{-4}]$), the considered non-coherent transmission architecture is comparable to the \ac{CE} performance of the conventional coherent \ac{MIMO}-\ac{OFDM} systems.

\subsection[]{\ac{AUD}}
\label{sec:Sim_AUD}

In this section, we evaluate the \ac{AUD} performance of the proposed \ac{BiGaBP} method.
Although the \ac{AUD} performance may be examined in terms of either \ac{FA}, \ac{MD}, or both, \ac{FA} can be removed at higher layers by leveraging cyclic redundancy check codes \cite{HaraAccess19}, which are widely employed in practice.
In light of the above, in this article, we adopt the occurrence of \acp{MD} as an \ac{AUD} performance index.

\begin{figure}[H]
        \centering
    \begin{subfigure}[b]{0.48\textwidth}
        \includegraphics[width=\textwidth]{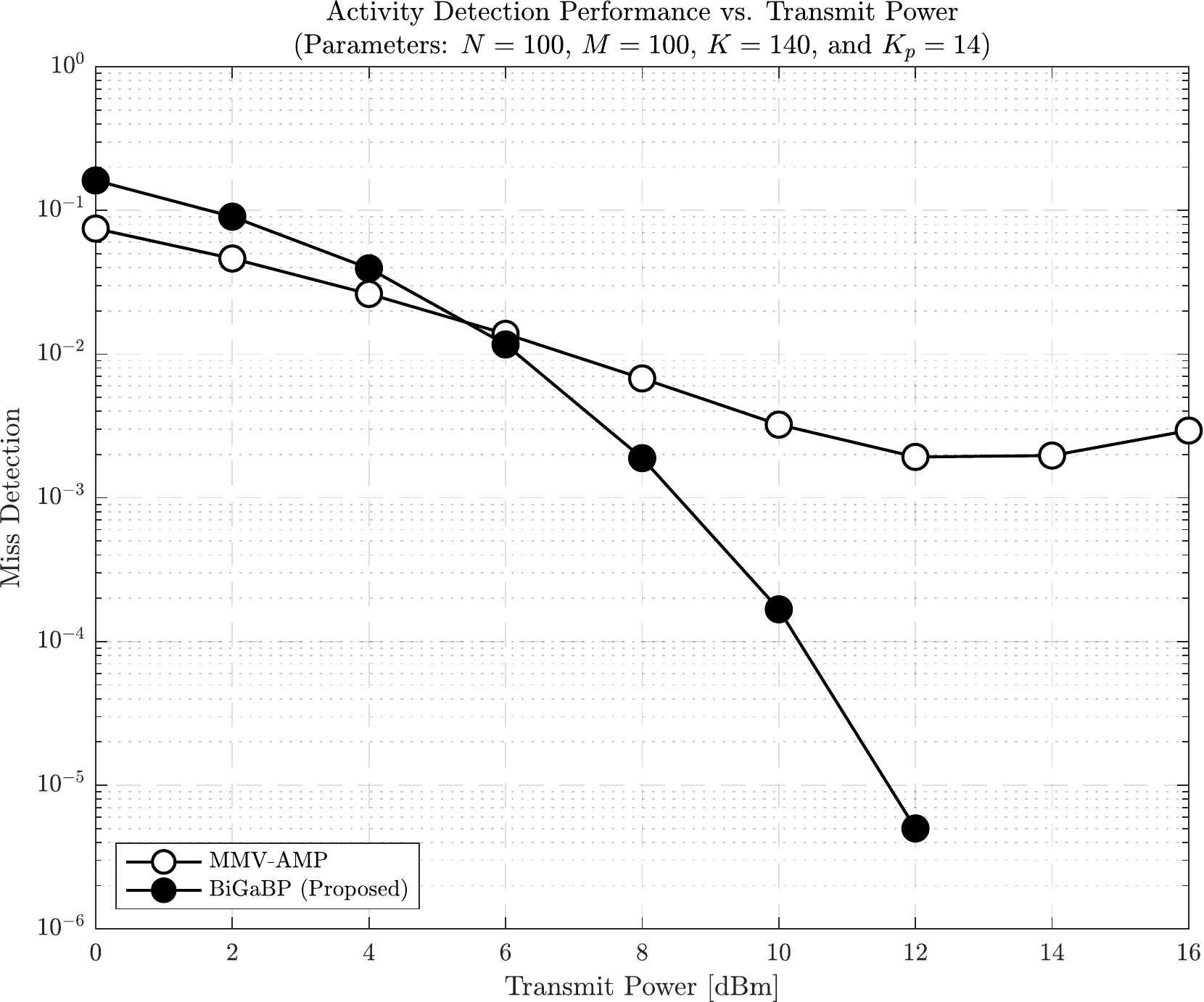}
        \caption{$15$ kHz subcarrier spacing ($N=M=100$ and $K_p=14$)}
        \label{fig:aer_15kHZ}
    \end{subfigure}
        ~ 
        \begin{subfigure}[b]{0.48\textwidth}
        \includegraphics[width=\textwidth]{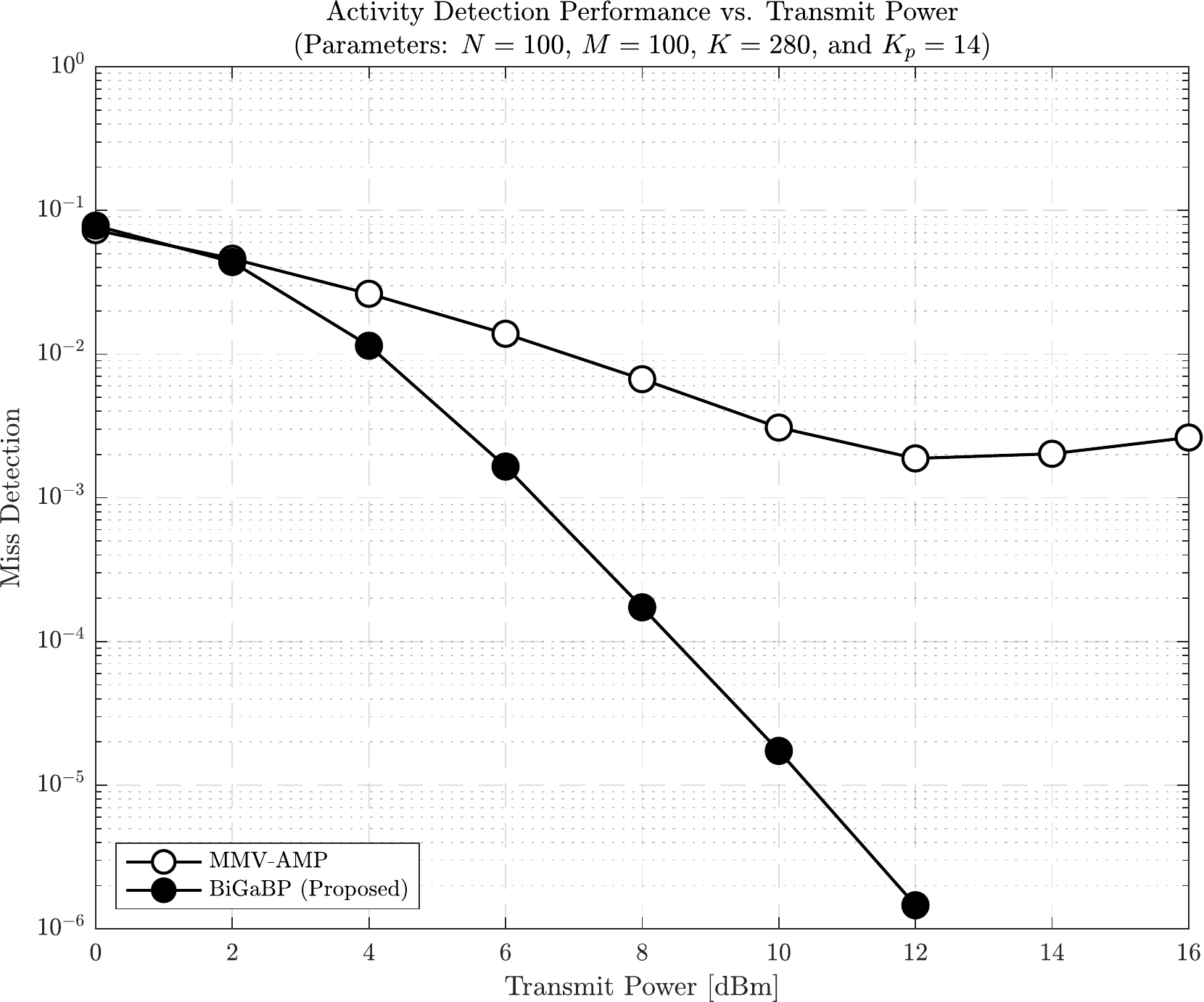}
     \caption{$30$ kHz subcarrier spacing ($N=M=100$ and $K_p=14$)}
      \label{fig:aer_30kHZ}
    \end{subfigure}
    \caption[]{\ac{MD} comparisons as a function of transmit power}
    \label{fig:aer}
    \end{figure}

In Figure \ref{fig:aer}, \ac{MD} probabilities of the proposed \ac{BiGaBP} and \ac{MMV-AMP} algorithms are illustrated for different symbol lengths $K$ as a function of transmit power at each uplink user, while assuming $K_p=14$ for both scenarios.
It is perceived from Figure \ref{fig:aer} that the proposed method can exponentially reduce the occurrence of \acp{MD} as transmit power increases, the reason of which can be explained from the discussions given in the preceding sections as follow.
As we observed from Figure \ref{fig:ber}--\ref{fig:nmse}, the proposed \ac{BiGaBP} algorithm starts to gradually recover the data and the channel from the observations $\bm{Y}$ as transmit power increases, which stems from the fact that the residual noise variances given in equation \eqref{eq:residualnoise_var_x} and \eqref{eq:residualnoise_var_h} are also accordingly reduced.
Consequently, the resultant LLR given in \eqref{eq:LLR} intends to be positive when $|\hat{h}_{nm}|$ is not sufficiently close to $0$ and negative when $\prod^N_{n=1}\mathcal{CN}\left(0, \gamma_{nm}+\psi^h_{nm}|\hat{h}_{nm}\right)\approx 0$ in comparison with $\prod^N_{n=1}\mathcal{CN}\left(0, \psi^h_{nm}|\hat{h}_{nm}\right)$ for a small $\psi^h_{nm}$.
Furthermore, the reason why the \ac{MD} performance of \ac{MMV-AMP} deteriorates in high transmit power regions can be explained as follows.

Besides the insufficient observations due to a non-orthogonal pilot structure, \ac{MMV-AMP} suffers from the fact that in such a high \ac{SNR} region, its estimation error noise variance becomes indistinguishable from the \ac{AWGN} noise level at the receiver, leading to a tendency to regard inactive users as active and vice versa.
In contrast, the proposed method mitigates this bottleneck by taking advantage of \acp{DoF} in the time domain.

\subsection{MSE Performance Prediction and Its Accuracy}

Finally, in this subsection we evaluate the accuracy of \ac{MSE} tracking via the state evolution of the proposed \ac{BiGaBP} algorithm\footnote{Note that since the \ac{GaBP} algorithm is a generalization of \ac{AMP}, the resultant error level can be predicted in a similar fashion to the state evolution in \ac{AMP}. For the sake of consistency, we call the corresponding error predicting quantities as \ac{BiGaBP}'s state evolution for the \ac{MSE} performances.}, where the predicted \ac{MSE} performances of the data and channel are obtained by equation \eqref{eq:residualnoise_var_x} and \eqref{eq:residualnoise_var_h}, respectively.

\begin{figure}[H]
\centering
\begin{subfigure}[b]{\columnwidth}
\includegraphics[width=\columnwidth]{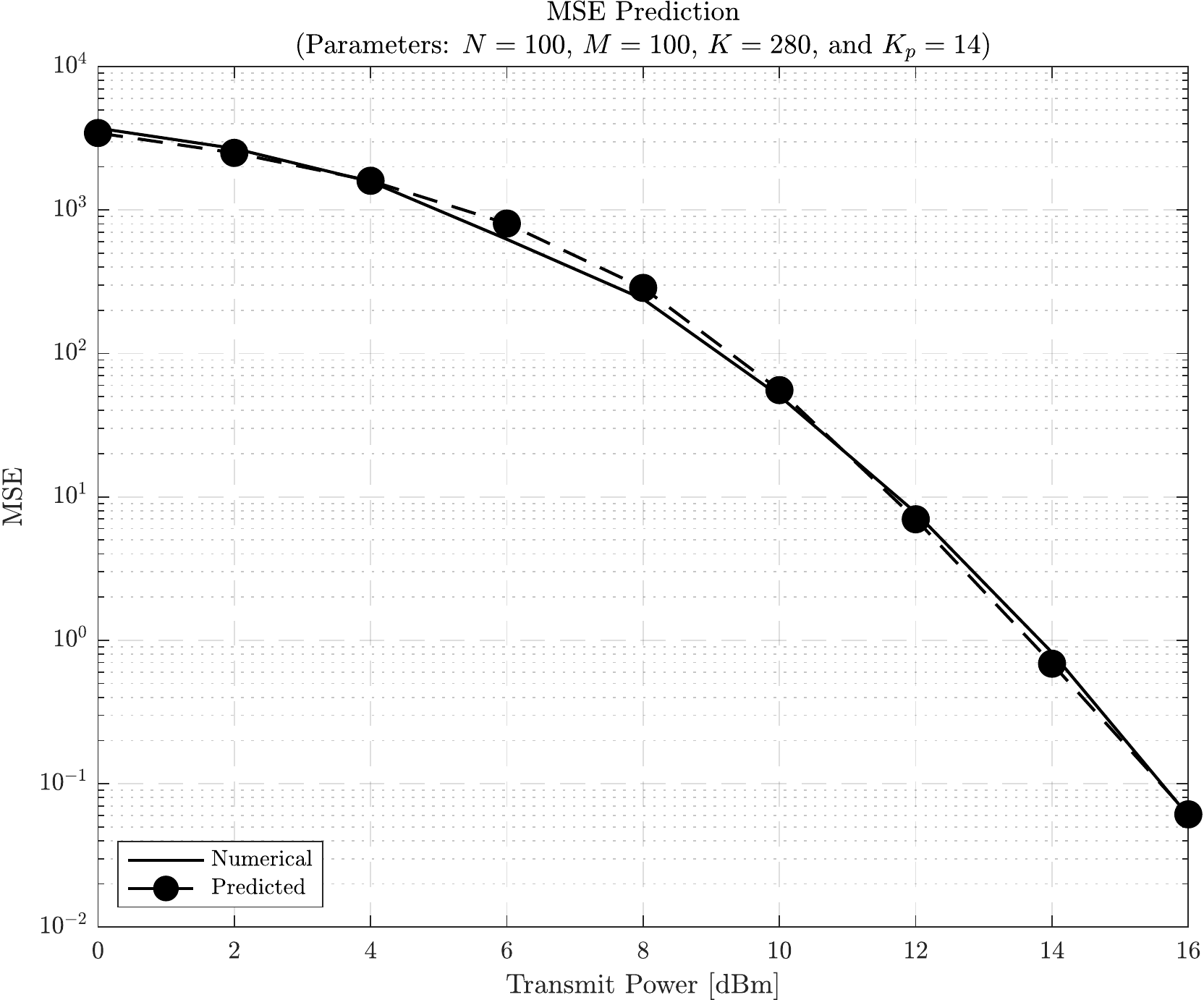}
\caption{\Ac{MSE} error of $\hat{\bm{X}}$}
\label{fig:MSE_Pre_X}
\vspace{1ex}
\end{subfigure}
\begin{subfigure}[b]{\columnwidth}
\includegraphics[width=\columnwidth]{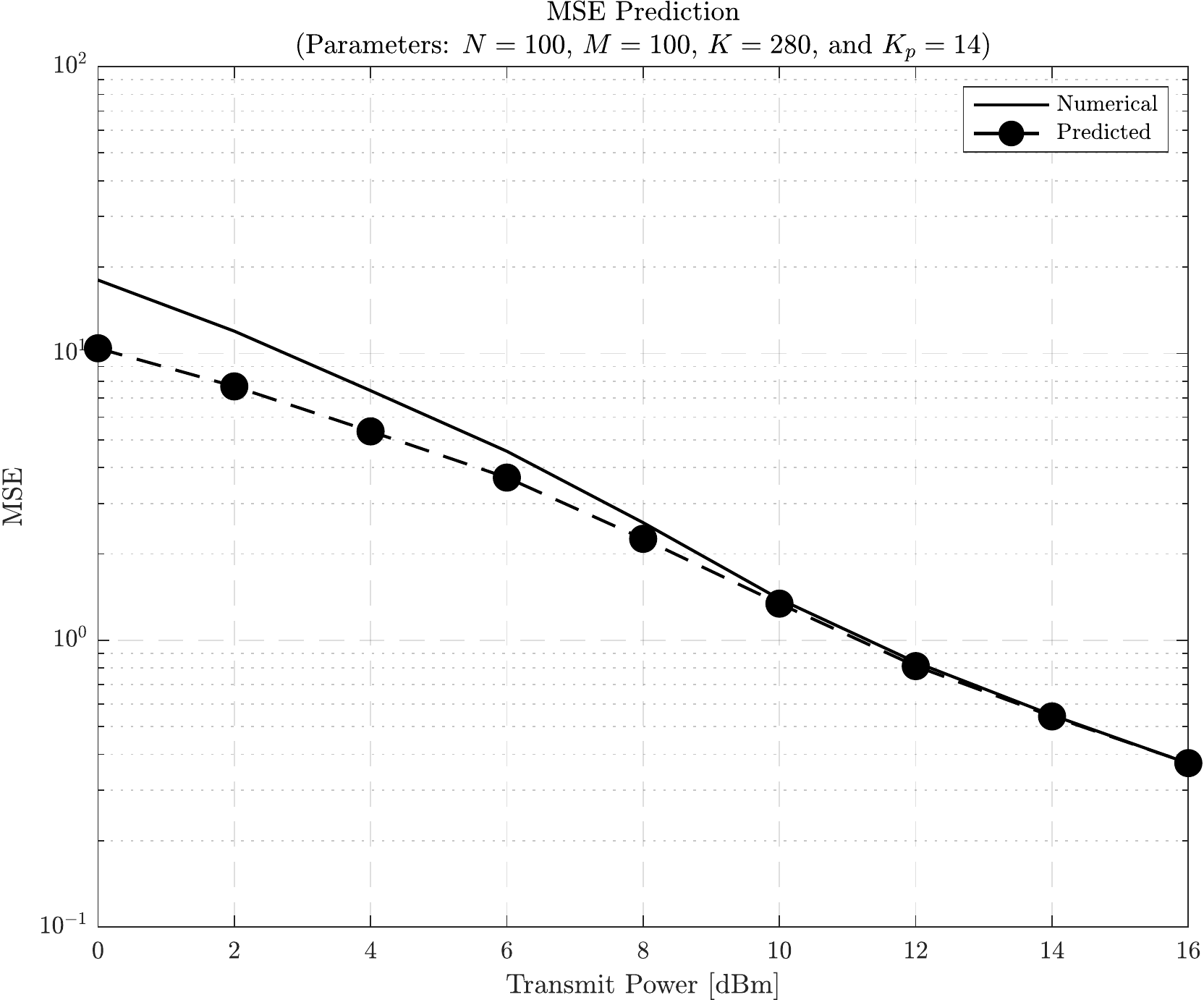}
\caption{\Ac{MSE} error of $\hat{\bm{H}}$}
\label{fig:MSE_Pre_H}
\end{subfigure}
\caption{MSE performance prediction via the state evolution of Algorithm \ref{alg:alg1} in comparison with the actual numerical evaluation as a function of transmit power.}
\label{fig:MSE_Prediction}
\vspace{-1ex}
\end{figure}

In Figure \ref{fig:MSE_Pre_X} and \ref{fig:MSE_Pre_H}, the predicted \ac{MSE} for $\hat{\bm{X}}$ and $\hat{\bm{H}}$ is, respectively, compared with its corresponding simulated counterpart, where the solid line and the dashed line with circle markers corresponds to the simulated and predicted performance, respectively.
Without loss of generality, the simulation setup is set to $K=280$ and $K_p=14$ due to the space limitation. 
It can be observed from the figures that the state evolution can track the error performance of the proposed \ac{BiGaBP} for both data and channel estimates, since the predicted \acp{MSE} follow approximately the same trajectory of its simulated counterpart.

\vspace{-1ex}
\section{Conclusion}

In this article, we proposed a novel \ac{JACDE} mechanism based on bilinear inference for \ac{OFDM} cell-free grant-free \ac{MIMO} systems \ul{without spreading data sequences}, while designing a severely non-orthogonal pilot sequence via frame theory with the aim of an efficient overhead reduction.

In order to sustain moderate throughput per user, in contrast with most of the grant-free literature, the proposed method is developed based on the conventional \ac{MIMO} \ac{OFDM} protocol, whereas employing activity detection capability without resorting to spreading informative data symbols, which is enabled by \ul{bilinear inference and pseudo-orthogonality of the independently-generated data symbols}.
To this end, we derived new Bayesian message passing rules based on Gaussian approximation, which enables \ac{JACDE}.
Feasibility of \ac{JACDE} grant-free access with \emph{non-spread} data sequences has been established via simulation-based performance assessment, which is expected to pose a new angle to related research topics.

\appendices
\section{Proof of Theorem \ref{theorem:QCSIDCO}}
\label{app:QCSIDCO}

Leveraging slack variables $t_{\ell,R}$ and $t_{\ell,I}$, the real and imaginary parts of $\tilde{\bm{F}}^{\rm H}_\ell \bm{f}_\ell$ can be bounded as
\begin{subequations}\label{eq:app_bound}
\begin{eqnarray}
\left|\Re\left\{\tilde{\bm{F}}^{\rm H}_\ell \bm{f}_\ell\right\}\right| &\leq& t_{\ell,R}\cdot \mathbf{1}_{L-1}\\
\left|\Im\left\{\tilde{\bm{F}}^{\rm H}_\ell \bm{f}_\ell\right\}\right| &\leq& t_{\ell,I}\cdot \mathbf{1}_{L-1},
\end{eqnarray}
\end{subequations}
which readily yields
\begin{equation}
\|\tilde{\bm{F}}^{\rm H}_\ell \bm{f}_\ell\|_\infty \leq \sqrt{t^2_{\ell,R}+t^2_{\ell,I}}.
\end{equation}

Furthermore, equation \eqref{eq:app_bound} can also be rewritten as
\begin{subequations}\label{eq:app_bound2}
\begin{eqnarray}
\footnotesize
\underbrace{\begin{cases}
 \Re\big\{\tilde{\bm{F}}_\ell\big\}^{\rm T}\Re\big\{\bm{f}_\ell\big\} + \Im\big\{\tilde{\bm{F}}_\ell\big\}^{\rm T}\Im\big\{\bm{f}_\ell\big\} - t_{\ell,R}\cdot \mathbf{1}_{L-1} \leq \mathbf{0}\\
-\big(\Re\big\{\tilde{\bm{F}}_\ell\big\}^{\rm T}\Re\big\{\bm{f}_\ell\big\} + \Im\big\{\tilde{\bm{F}}_\ell\big\}^{\rm T}\Im\big\{\bm{f}_\ell\big\}\big) - t_{\ell,R}\cdot \mathbf{1}_{L-1} \leq \mathbf{0} \end{cases}
}_{\Leftrightarrow \left|\Re\left\{\tilde{\bm{F}}^{\rm H}_\ell \bm{f}_\ell\right\}\right| - t_{\ell,R}\cdot \mathbf{1}_{L-1}\leq \mathbf{0}}\!\!\!\! \\
\footnotesize
\underbrace{\begin{cases}
 \Re\big\{\tilde{\bm{F}}_\ell\big\}^{\rm T}\Im\big\{\bm{f}_\ell\big\} - \Im\big\{\tilde{\bm{F}}_\ell\big\}^{\rm T}\Re\big\{\bm{f}_\ell\big\} - t_{\ell,I}\cdot \mathbf{1}_{L-1} \leq \mathbf{0}\\
-\big(\Re\big\{\tilde{\bm{F}}_\ell\big\}^{\rm T}\Im\big\{\bm{f}_\ell\big\} - \Im\big\{\tilde{\bm{F}}_\ell\big\}^{\rm T}\Re\big\{\bm{f}_\ell\big\}\big) - t_{\ell,I}\cdot \mathbf{1}_{L-1} \leq \mathbf{0} \end{cases}
}_{\Leftrightarrow \left|\Im\left\{\tilde{\bm{F}}^{\rm H}_\ell \bm{f}_\ell\right\}\right| - t_{\ell,I}\cdot \mathbf{1}_{L-1}\leq \mathbf{0}} \!\!\!\!
\end{eqnarray}
\end{subequations}
where the inequality is applied in an element-by-element manner, leading to 
\begin{subequations}
\begin{eqnarray}
\small
\underbrace{\begin{bmatrix}
\Re\big\{\tilde{\bm{F}}_\ell\big\}^{\rm T}&\!\!\!\!
\Im\big\{\tilde{\bm{F}}_\ell\big\}^{\rm T}&\!\!\!\!
-\mathbf{1}_{(L-1)\times 1}&\!\!\!\!
\mathbf{0}_{(L-1)\times 1}
\end{bmatrix}}_{\triangleq \bm{A}_{\ell,R,1}}
\bm{x}_\ell \leq& \mathbf{0}\nonumber\\[-3.2ex]\\
\small
\underbrace{\begin{bmatrix}
-\Re\big\{\tilde{\bm{F}}_\ell\big\}^{\rm T}&\!\!\!\!
-\Im\big\{\tilde{\bm{F}}_\ell\big\}^{\rm T}&\!\!\!\!
-\mathbf{1}_{(L-1)\times 1}&\!\!\!\!
\mathbf{0}_{(L-1)\times 1}
\end{bmatrix}}_{\triangleq \bm{A}_{\ell,R,2}}
\bm{x}_\ell \leq& \mathbf{0}\nonumber\\[-3.2ex]\\
\small
\underbrace{\begin{bmatrix}
-\Im\big\{\tilde{\bm{F}}_\ell\big\}^{\rm T}&\!\!\!\!
\Re\big\{\tilde{\bm{F}}_\ell\big\}^{\rm T}&\!\!\!\!
\mathbf{0}_{(L-1)\times 1}&\!\!\!\!
-\mathbf{1}_{(L-1)\times 1}
\end{bmatrix}}_{\triangleq \bm{A}_{\ell,I,1}}
\bm{x}_\ell \leq& \mathbf{0}\nonumber\\[-3.2ex]\\
\small
\underbrace{\begin{bmatrix}
\Im\big\{\tilde{\bm{F}}_\ell\big\}^{\rm T}&\!\!\!\!
-\Re\big\{\tilde{\bm{F}}_\ell\big\}^{\rm T}&\!\!\!\!
\mathbf{0}_{(L-1)\times 1}&\!\!\!\!
-\mathbf{1}_{(L-1)\times 1}
\end{bmatrix}}_{\triangleq \bm{A}_{\ell,I,2}}
\bm{x}_\ell \leq& \mathbf{0}\nonumber\\[-3.2ex]
\end{eqnarray}
\end{subequations}
where $\bm{x}_\ell$ is defined in Theorem \ref{theorem:QCSIDCO}, equation \eqref{eq:T_ball_const} can be readily obtained from equation \eqref{eq:T_ball_const_original}, and this completes the proof.

\section{Derivation of equation \eqref{eq:effective_PDF} and \eqref{eq:normalizing}}
\label{app:effective_PDF}

Given equation \eqref{eq:extrinsic_PDF_h} and \eqref{eq:expected_h_mean}, the effective \ac{PDF} can be readily expressed as
\begin{align}
&p_{\bm{l}^h_{k,m}|\bm{h}_{m}}(\bm{l}^h_{k,m}|\bm{h}_{m})p_{\bm{h}_{m}}(\bm{h}_{m})\nonumber\\
&= p_{\bm{h}_{m}}(\bm{h}_{m}) \mathcal{CN}_N\left(\bm{\mu}^h_{k,m},\bm{\Sigma}^h_{k,m}\right)\nonumber\\
&= \left[ \lambda\:\mathcal{CN}_{\!N}\left(0, \mathbf{\Gamma}_m\right) + (1-\lambda)\delta(\bm{h}_{m})\right] \mathcal{CN}_N\left(\bm{\mu}^h_{k,m},\bm{\Sigma}^h_{k,m}\right)\nonumber\\
&= \Bigg[ \frac{\lambda\exp\big( \!- {\bm{\mu}^h_{k,m}}^{\!\!\!\!\!\!\mathrm{H}}\left(\bm{\Sigma}^h_{k,m} + \mathbf{\Gamma}_m\right)^{-1}{\bm{\mu}^h_{k,m}}\big)}{\pi^N  |\mathbf{\Gamma}_m+\bm{\Sigma}^h_{k,m}|}\nonumber\\
&\cdot\mathcal{CN}_{\!N}\big(\!\big({\bm{\Sigma}^h_{k,m}}^{\hspace{-2ex}-1} + \mathbf{\Gamma}_m^{-1}\big)^{\!-1}{\bm{\Sigma}^h_{k,m}}^{\hspace{-2ex}-1}\bm{\mu}^h_{k,m},\big({\bm{\Sigma}^h_{k,m}}^{\hspace{-2ex}-1} + \mathbf{\Gamma}_m^{-1}\big)^{\!\!-1}\big)\nonumber\\
&\quad +  \frac{(1-\lambda)\exp\big(- {\bm{\mu}^h_{k,m}}^{\!\!\!\!\!\!\mathrm{H}}{\bm{\Sigma}^h_{k,m}}^{\hspace{-2ex}-1}{\bm{\mu}^h_{k,m}}\big)}{\pi^N |\bm{\Sigma}^h_{k,m} |}\delta(\bm{h}_{m})\Bigg]
\label{eq:effective_PDF_1}
\end{align}
where by the Woodbury inverse lemma we utilized 
\begin{equation}
\big({\bm{\Sigma}^h_{k,m}}^{\hspace{-2ex}-1} - {\bm{\Sigma}^h_{k,m}}^{\hspace{-2ex}-1}\big({\bm{\Sigma}^h_{k,m}}^{\hspace{-2ex}-1} \!+ \mathbf{\Gamma}_m^{-1}\big)^{\!-1} {\bm{\Sigma}^h_{k,m}}^{\hspace{-2ex}-1}\big) \!=\!  \left(\bm{\Sigma}^h_{k,m} \!+\! \mathbf{\Gamma}_m\right)^{\!-1}\!\!\!\!\!.
\end{equation}

Recalling $\left(\bm{A}^{-1}+\bm{B}^{-1}\right)^{-1} = \bm{B}\left(\bm{A}+\bm{B}\right)^{-1}\bm{A}$ for inversible $\bm{A}$ and $\bm{B}$, one may readily obtain equation \eqref{eq:effective_PDF} from \eqref{eq:effective_PDF_1}.
This completes the derivation of \eqref{eq:effective_PDF}.

Similarly, the normalizing factor $C_{k,m}$ is given by
\begin{align}
C_{k,m}& \!\triangleq\!\int_{\bm{h}^\prime_{m}}p_{\bm{l}^h_{k,m}|\bm{h}^\prime_{m}}(\bm{l}^h_{k,m}|\bm{h}^\prime_{m})p_{\bm{h}_{m}}(\bm{h}^\prime_{m}) \\
& = \frac{\lambda\exp\big( \!- {\bm{\mu}^h_{k,m}}^{\!\!\!\!\!\!\mathrm{H}}\left(\bm{\Sigma}^h_{k,m} + \mathbf{\Gamma}_m\right)^{-1}{\bm{\mu}^h_{k,m}}\big)}{\pi^N  |\mathbf{\Gamma}_m+\bm{\Sigma}^h_{k,m}|}\nonumber\\
&\quad +  \frac{(1-\lambda)\exp\big(- {\bm{\mu}^h_{k,m}}^{\!\!\!\!\!\!\mathrm{H}}{\bm{\Sigma}^h_{k,m}}^{\hspace{-2ex}-1}{\bm{\mu}^h_{k,m}}\big)}{\pi^N |\bm{\Sigma}^h_{k,m} |}\delta(\bm{h}_{m})\nonumber\\
&= \frac{\lambda\exp\big( \!- {\bm{\mu}^h_{k,m}}^{\!\!\!\!\!\!\mathrm{H}}\left(\bm{\Sigma}^h_{k,m} + \mathbf{\Gamma}_m\right)^{-1}{\bm{\mu}^h_{k,m}}\big)}{\pi^N  |\mathbf{\Gamma}_m+\bm{\Sigma}^h_{k,m}|}\nonumber\\
& \cdot\Big(1 + \frac{1-\lambda}{\lambda} 
 \big|{\bm{\Sigma}^h_{k,m}}^{\hspace{-2ex}-1}\mathbf{\Gamma}_m+\mathbf{I}_{N} \big|\exp\big(-\pi^h_{k,m}\big) \Big)\nonumber,
\end{align}
which completes the derivation of equation \eqref{eq:normalizing}.

\bibliographystyle{IEEEtran}
\bibliography{listofpublications}

\end{document}